\documentclass[aps,prb,twocolumn,showpacs,amsmath,amssymb,superscriptaddress,floatfix]{revtex4-1}
\usepackage{graphicx}
\usepackage{dcolumn} 
\usepackage{bm} 
\usepackage{amsfonts}
\usepackage{amsmath}
\usepackage{ulem}
\usepackage{color}
\usepackage{hyperref}
\usepackage{ulem}

\usepackage{subfiles}

\newcommand{\Vector}[1]{\ensuremath{\left( \begin{array}{c} #1 \end{array} \right)}}
 
\begin{document}

\title{The Dirac paradox in 1+1 dimensions  \\and its realization with spin-orbit coupled nanowires} 

\author{Leonid Gogin}
\affiliation{Dipartimento di Scienza Applicata e Tecnologia del Politecnico di Torino, I-10129 Torino, Italy}

\author{Lorenzo Rossi}
\email{lorenzo.rossi@polito.it}
\affiliation{Dipartimento di Scienza Applicata e Tecnologia del Politecnico di Torino, I-10129 Torino, Italy}

\author{Fausto Rossi}
\affiliation{Dipartimento di Scienza Applicata e Tecnologia del Politecnico di Torino, I-10129 Torino, Italy}

\author{Fabrizio Dolcini}

\affiliation{Dipartimento di Scienza Applicata e Tecnologia del Politecnico di Torino, I-10129 Torino, Italy}

\begin{abstract}
At the interface   between two massless Dirac models with opposite helicity a paradoxical situation arises: A transversally  impinging electron can seemingly neither be transmitted nor reflected, due to the locking between spin and momentum.  Here we investigate this paradox in one  spatial dimension where, differently from higher dimensional realizations, electrons cannot leak along the interface. We  show that  models involving only massless Dirac modes lead to either no solutions or to trivial solutions to the paradox, depending on how the helicity change across the interface is modeled. However,  non trivial scattering solutions to the paradox are shown to exist when additional massive Dirac modes are taken into account. Although these modes carry no current for energies within their gap,  their interface coupling with the massless modes can  induce a finite and tunable transmission. 
Finally, we show that such  massless+massive Dirac model can be realized in suitably gated spin-orbit coupled nanowires exposed to an external Zeeman field, where the transmission coefficient can be controlled electrically.
 
\end{abstract}

\maketitle
\section{Introduction}
Conventional semiconductor heterostructures are typically described, within the envelope function and effective mass approximations,   by a Schr\"odinger  Hamiltonian with a space dependent effective electron mass  varying along the growth direction and accounting for the different effective masses of the component materials.

\begin{figure}[h]
\centering
\includegraphics[width=0.9\linewidth]{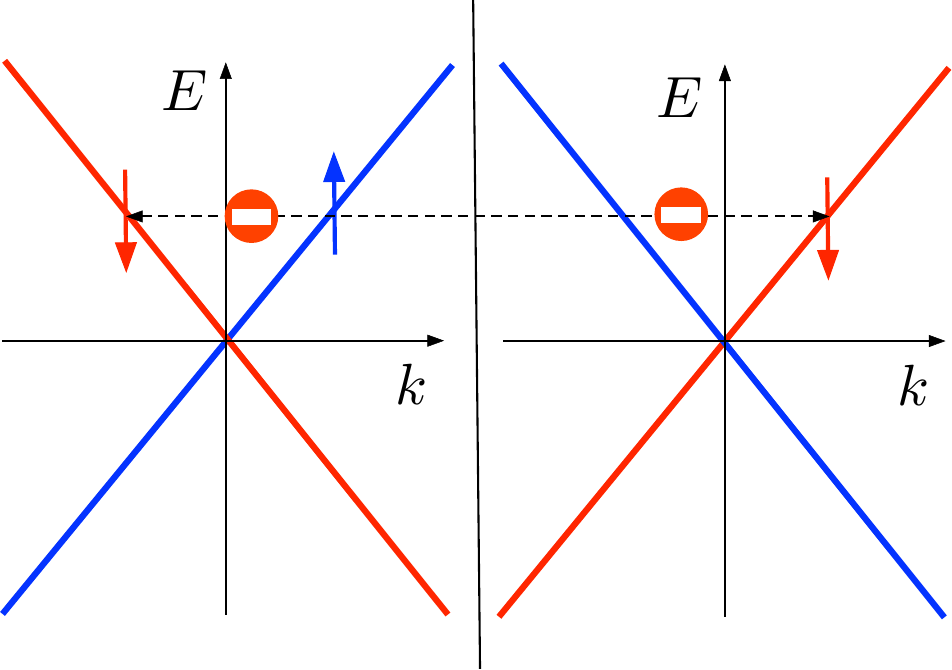}
\caption{\label{Fig1} Schematic representation of the Dirac paradox emerging at a heterojunction between two massless Dirac models with opposite helicity. Blue and red lines describe spin-$\uparrow$ and spin-$\downarrow$ states, respectively. On the left-hand side of the junction, right-moving electrons are characterized by spin-$\uparrow$ and left-moving electrons by spin-$\downarrow$,   while the opposite occurs on the right-hand side. A spin-$\uparrow$ electron impinging    from the left onto the interface can seemingly neither be transmitted nor reflected, due to spin conservation. }
\end{figure}

In the last two decades, however, it has been realized that in  various materials such as graphene, topological insulators and Weyl semimetals,  the dynamics of the conduction electrons is well captured,   in physically relevant regimes,    by a (D+1)-dimensional {\it massless} Dirac electron model\cite{geim-novoselov,castroneto-review,dassarma_review_2011,kane-hasan-review,zhang-review,hasan_natmat_2016,yang_2016}, where~D denotes the spatial dimension, and ``$+1$" the time dimension. These discoveries have thus spurred   the interest in the investigation of Dirac   heterojunctions. Each massless Dirac cone is characterized by a given helicity of the electron eigenstates, i.e. a sign encoding the locking between the propagation direction  and the orientation of a ``spin-like" degree of freedom, which can be a sublattice pseudospin, like in graphene, or the actual angular momentum in topological insulators. In particular, when a junction is formed between two Dirac materials with opposite helicity, a paradoxical situation emerges, as sketched  in Fig.\ref{Fig1}. A right-moving electron (blue line on the left-hand side) impinging transversally towards a spin-inactive interface    can neither be transmitted   nor be reflected, due to spin conservation. 
The Dirac paradox has been discussed in heterojunctions between two 3D Topological insulators, whose surface states are governed by a  2D massless Dirac Hamiltonian. In such a case the surface  electrons turn out to ``escape" the problem by leaking along the interface surface\cite{murakami_2011,sen_2012,debeule_2013,asano_2013}.  However, in a  truly 1D realization of a Dirac model such way out to an extra dimension does not exist   and the Dirac paradox becomes even more interesting. The challenging question  is whether  a solution in 1D exists and, if so, whether it can be realized in some  physical system.\\
In this paper we investigate the Dirac paradox in 1D and address these problems. First, we show that, if the helicity change across the interface is accounted for by an inhomogeneous velocity profile, the paradox has no solution, in the sense that the continuity equation forbids the existence of scattering states and only allows for eigenstates that involve electron injection from both sides of the junction, which  carry a vanishing current. If, however,   the interface directly introduces spin-rotation processes, the solution of the paradox is of course trivial and transmission is possible.  We then investigate whether non-trivial solutions with proper scattering states and finite transmission exist   without  a spin-active interface.  
To this purpose, we propose an extended model   involving both massless and massive Dirac modes. We show that, despite carrying no current for energies within their gap,  the massive modes  play a crucial role  in inhomogeneous problems like the Dirac paradox.  In particular,  as will be discussed in details, scattering states describing a transmission from a spin-$\uparrow$ incoming massless mode to a spin-$\downarrow$ outgoing massless mode do exist, due to the interface coupling between  massless and massive modes. Moreover, the resulting transmission coefficient is finite and  tuneable.

We then discuss the possible realization of such extended model. While massless Dirac helical states have been proven to exist  at the edges of   quantum spin Hall systems\cite{bernevig_science_2006,zhang_2008,molenkamp-zhang,knez_2011,knez_2015}, this implementation is not optimal for the Dirac paradox in 1D. Indeed, since these states flow  at the boundaries of a 2D quantum well, an heterojunction between two such wells with opposite edge helicity would exhibit a linear interface, whereto electrons could leak, like in the case of   heterojunctions between two 3D Topological insulators mentioned above. However,  a truly 1D implementation of  helical  states has been realized with spin-orbit coupled nanowires (NWs) exposed to a magnetic field\cite{depicciotto_2010,kouwenhoven_nanolett_2013,kouwenhoven_natcom_2017,kouwenhoven_2012,liu_2012,heiblum_2012,xu_2012,defranceschi_2014,marcus_2016,marcus_science_2016,kouwenhoven_2018}, in the regime of spin-orbit energy   much larger than the Zeeman energy\cite{streda,vonoppen_2010,dassarma_2010,loss_PRB_2011,lutchyn_2012,loss_PRB_2017}. 

So far, this remarkable discovery  has been    mostly exploited in the search for Majorana quasi-particles\cite{vonoppen_2010,dassarma_2010,kouwenhoven_2012,liu_2012,heiblum_2012,xu_2012,defranceschi_2014,marcus_2016,marcus_science_2016,kouwenhoven_2018,kitaev,alicea_review,fujimoto,aguado_review}. However,   further interesting research areas  on NWs are fostered by  the recent  advances in gating techniques\cite{sasaki_2013,micolich,sasaki_2017,das_2019,guo_2021,sasaki_2021}, which nowadays enable one to  control  the Rashba spin-orbit coupling (RSOC), both in magnitude\cite{gao_2012,slomski_NJP_2013,wimmer_2015,nygaard_2016,sherman_2016,tokatly_PRB_2017,loss_2018,goldoni_2018,tsai_2018,gao-review,lau_2021} and sign\cite{kaindl_2005,slomski_NJP_2013,wang-fu_2016,nitta-frustaglia}. Because in a NW the helicity  of the massless modes is determined by the sign of the RSOC, a NW with   two differently gated regions can  represent  a truly 1D implementation of the Dirac paradox configuration. Notably, in such an inhomogeneous setup, the  massless helical modes are not sufficient to describe the low energy physics, which turns out to be well captured by the  massless+massive Dirac model we propose  here, instead. The resulting conductance can  be tuned electrically over a wide range of values.

Our paper is organized as follows. In Sec.\ref{sec-2} we analyze the Dirac paradox with two different models involving only massless modes. Then, in Sec.\ref{sec-3} we introduce a model with both massless and massive Dirac modes and show how this can yield a finite transmission coefficient depending on three parameters. Furthermore, in Sec.\ref{sec-4} we show that this model can be implemented in a suitably designed setup involving spin-orbit coupled NWs. Finally, in Sec.\ref{sec-5} we discuss our results and draw our conclusions.

\section{Massless Dirac heterojunctions}
\label{sec-2}
Let us thus consider a junction connecting two 1+1 dimensional massless Dirac models 
\begin{equation}
\hat{\mathcal{H}}_{L/R}=  v_{L/R}  \int \hat{\Psi}^\dagger(x)\sigma_z   p_x \, \hat{\Psi}^{}(x)\, dx
\end{equation}
where $\hat{\mathcal{H}}_{L}$and $\hat{\mathcal{H}}_{R}$ denote the Hamiltonians  on the left and on the right side of the interface region, respectively, $\hat{\Psi}=(\hat{\Psi}_\uparrow,\hat{\Psi}_\downarrow)^T$ is the $2\times 1$ electron spinor field operator, $p_x=-i \hbar \partial_x$ is the  momentum operator, and $\sigma_z$ is a Pauli matrix in spin space. Finally $v_{L/R}$ denotes the Fermi velocity. When $v_R$ and $v_L$ have opposite signs, the   helicity changes across the interface  and the Dirac paradox emerges. The answer to the paradox, if any, heavily depends on how the crossover from $\hat{\mathcal{H}}_{L}$ to  $\hat{\mathcal{H}}_{R}$ occurs, as we shall discuss here below considering different models.

\subsection{Model 1:  velocity sign change}
The most straightforward way to implement the crossover from $\hat{\mathcal{H}}_{L}$ to  $\hat{\mathcal{H}}_{R}$ is to assume that the  entire system is characterized by an inhomogeneous velocity $v(x)$, varying from  $v_L$ to  $v_R$ over a certain crossover length $\lambda$.
Since  the  momentum operator $p_x$  does not commute with an inhomogeneous velocity profile $v(x)$, a quite natural approach is to replace their product $p_x\, v$ by a half of their anticommutator, obtaining  the following Hamiltonian
\begin{equation}\label{Ham-model-1}
\hat{\mathcal{H}}=\int \hat{\Psi}^\dagger(x)\sigma_z  \frac{\left\{ v(x)\,, p_x\right\}}{2} \, \hat{\Psi}^{}(x)\, dx\quad.
\end{equation} 
The current operator associated to Eq.(\ref{Ham-model-1}) is
\begin{equation}\label{curr-model-1}
\hat{J}(x)={\rm e}\, v(x)\hat{\Psi}^\dagger(x) \sigma_z \hat{\Psi}^{}(x) \quad,
\end{equation}
with ${\rm e}$ denoting the electron charge,  whereas the  Heisenberg Equation dictated by Hamiltonian (\ref{Ham-model-1}) reads
\begin{equation}
\partial_t  \hat{\Psi} =-\sigma_z \left( v(x) \partial_x \hat{\Psi}+\frac{\partial_x v}{2} \hat{\Psi}\right) \quad.\label{HE}
\end{equation}
Looking for stationary solutions $\hat{\Psi}(x,t)=\hat{\Psi}_E(x) e^{-i E t/\hbar}$ and multiplying Eq.(\ref{HE}) by $\sigma_z$ on the left, the equation reduces to
\begin{equation}
\partial_x \hat{\Psi} = v^{-1}(x) \left(  - \frac{\partial_x v}{2} \sigma_0 + i \frac{E}{\hbar} \sigma_z  \right)   \hat{\Psi} \quad,
\end{equation}
whose formal solution is
\begin{eqnarray}
\hat{\Psi}_E(x)&=&\displaystyle \exp\left[-\frac{1}{2} \int_{x_R}^x \frac{\partial_x v}{v(x^\prime)} dx^\prime\right] \times \nonumber \\
& & \times \exp\left[iE \sigma_z \int_{x_R}^x \frac{dx^\prime}{\hbar v(x^\prime)} \right]\,\hat{\Psi}_E(x_R)  \quad, \label{sol-pre}
\end{eqnarray}
where $x_R$ is some arbitrary reference space point.
One can now exploit $\partial_x \ln |v(x)|= \partial_x v/v(x)$,    denote $k_E(x)=E/\hbar v(x)$ and  write
\begin{equation}
\hat{\Psi}_E(x_R) = \frac{{u}}{\sqrt{2\pi \hbar |v(x_R)|}}  \, \hat{a}_{E}\quad,
\end{equation}
where ${u}$ is   a position-independent $2\times 1$ spinor and $\hat{a}_E$  the related energy-$E$ mode operator fulfilling $\{ a^{}_{E} \,, a^{\dagger}_{E^\prime}\} =\delta(E-E^\prime)$. At each energy $E$ there are thus two independent solutions, corresponding to two mutually orthogonal choices for the spinor $u$.  Then,   Eq.(\ref{sol-pre}) takes the form
\begin{eqnarray}
\hat{\Psi}_E(x)&=&\displaystyle   \frac{1}{\sqrt{2\pi \hbar|v(x)|}} \, e^{i\sigma_z \int_{x_R}^x k_E(x^\prime) dx^\prime }\, u \, \hat{a}_E \quad, \label{sol}
\end{eqnarray}
which straightforwardly implies   that at any space point~$x$, including possible discontinuity points of $v(x)$, the following boundary condition holds
\begin{equation}\label{continuity}
\sqrt{|v(x^+)|} \hat{\Psi}_E(x^+) =\sqrt{|v(x^-)|} \hat{\Psi}_E(x^-)
\end{equation}
where $x^\pm=x\pm\varepsilon$ with $\varepsilon \rightarrow 0$.

If $v(x)$ varies in magnitude  from $v_L$ to $v_R$ while  preserving   a (say) positive sign, $v(x)=|v(x)|$, the Hamiltonian~(\ref{Ham-model-1}) can equivalently be rewritten as 
\begin{equation}\label{Ham-inhomo-2}
\hat{\mathcal{H}}=\int \hat{\Psi}^\dagger(x)  \sigma_z \sqrt{v(x)} \left[ p_x \sqrt{v(x)} \right]\, \hat{\Psi}^{}(x)\, dx\quad.
\end{equation}
In this case, one finds that the transmission coefficient is always 1, regardless of the specific values of $v_L,v_R>0$, as discussed in Ref.[\onlinecite{peres}].

If, however, $v(x)$ vanishes at some point $x_0$,  like in the Dirac paradox, the problem becomes more subtle. Indeed in such case the energy dependent phase factor involving $k_E(x)$ in the solution Eq.(\ref{sol})  is well defined only if $v(x)$ vanishes as $|v(x) |=O(|x-x_0|^\alpha)$ with  $0<\alpha<1$. Moreover the solution diverges as $\sim 1/\sqrt{|v(x)|}$ for $x\rightarrow x_0$. Yet, in view of the condition (\ref{continuity}), the current in Eq.(\ref{curr-model-1}) is finite. Denoting by $\hat{J}_E^\pm \doteq \hat{J}_E(x_0^\pm)$  the current operator   for a stationary solution at energy $E$ at the two sides of the point $x_0$ of vanishing velocity, one straightforwardly finds from Eq.(\ref{continuity}) that $\hat{J}_E^+=-\hat{J}_E^-$.
  For stationary solutions, however, the continuity equation requires  the expectation value of the current to be continuous and  independent of the position. The only possibility is that  no current flows through the system, $\langle  \hat{J}_E(x)\rangle \equiv 0\, \, \forall x$, implying that  the   spinor   $u$ appearing in Eq.(\ref{sol}) {\it must be} chosen to have vanishing spin along~$z$, i.e. $  u^\dagger  \sigma_z u  =0$. Up to an overall  dimensional coefficient, two independent choices are   $u_{+}  = (1\,, e^{i \phi})^T/\sqrt{2}$ and $u_{-} = (e^{-i \phi},-1)^T/\sqrt{2}$,  where $\phi$ is an arbitrary phase.  \\

As an illustrative example, consider for instance the spatially odd profile $v(x)=- v_F\,\mbox{sgn}(x) \tanh^\alpha(|x|/\lambda)$, which describes a velocity sign change from $v_L=+v_F$ to $v_R=-v_F$ across the interface located at $x_0=0$, occurring over a lengthscale $\lambda$ and with an exponent $0<\alpha<1$. It is   straightforward to prove that the solution (\ref{sol}) is spatially even. Explicitly,  choosing e.g. $x_R=-4 \lambda$ as a reference point and   taking the phase $\phi=0$ in the above spinors $u$, one finds $k_E(x)\simeq -E \,\mbox{sgn}(x)/\hbar v_F$  for $|x|\gg \lambda$. The two physically correct solutions of the Heisenberg Eq.(\ref{HE}) then read
\begin{equation}
\hat{\Psi}_{\pm} (x,t)=  \int \frac{dE}{{\sqrt{2\pi \hbar v_F}}}\, \psi_{E \pm}(x)\, e^{-i E t/\hbar} \, \hat{a}_{E\pm}  \quad,
\end{equation}
where the wavefunctions  $\psi_{E \pm}$ for $|x| \gg \lambda$ take the form
\begin{equation}\label{asymp-waves}
\psi_{E\pm} (x)=\frac{e^{-i E |x|/\hbar v_F} }{\sqrt{2}} \left( \begin{array}{c}1 \\ 0 \end{array} \right) \pm \frac{e^{+i E |x|/\hbar v_F} }{\sqrt{2}} \left( \begin{array}{c}0 \\ 1 \end{array} \right) \quad.
\end{equation}
These solutions fulfill the continuity equation by carrying a vanishing current [see Eq.(\ref{curr-model-1})]. Note that the spatially even wavefunctions in Eq.(\ref{asymp-waves})  involve incoming waves from {\it both} sides and cannot be   scattering state solutions. Moreover, any attempt to construct scattering states by  their linear combinations would fail and would also violate the continuity equation. \\

In summary, the answer to the Dirac paradox provided by model 1 is that, when $v_L$ and $v_R$ have opposite signs, it is impossible to construct scattering state solutions that  respect the continuity equation. The transmission coefficient cannot be properly defined. Physically correct solutions must necessarily involve incoming waves from both sides and carry no current, regardless of the specific magnitudes of $|v_L|$ and $|v_R|$.
We conclude this section by noticing that the model 1  only involves the $\sigma_z$-component of spin [see Eq.(\ref{Ham-model-1})], and the space-dependent $v(x)$ changes magnitude and sign of such component. In this respect, the model is purely scalar.
\subsection{Model 2: spin-active interface}
The second model to approach the Dirac paradox is described by the Hamiltonian
\begin{equation}\label{Ham-model-2}
\hat{\mathcal{H}}=v_F \int \hat{\Psi}^\dagger(x) \left( e^{-i \theta(x) \sigma_x/2}    p_x \sigma_z  \, e^{+i \theta(x) \sigma_x/2}\right) \hat{\Psi}^{}(x)\, dx
\end{equation}
where the helicity changes sign through a counter-clockwise rotation of the $\sigma_z$ spin  around the $x$-axis by a    space-dependent angle $\theta(x)$  varying from $\theta_L=0$ to $\theta_R=\pi$, over a certain crossover length. Thus, differently from the purely scalar model (\ref{Ham-model-1}), the model (\ref{Ham-model-2}) exploits the full ${\rm SU}(2)$ spin structure and   Hamiltonian terms at two different points do not commute in general. The current operator related to the Hamiltonian (\ref{Ham-model-2}) is
\begin{eqnarray}\label{curr-model-2}
\hat{J}(x)&=& {\rm e} v_F \hat{\Psi}^\dagger(x)  \left( e^{-i \theta(x) \sigma_x/2}     \sigma_z  \, e^{+i \theta(x) \sigma_x/2}\right) \hat{\Psi}(x)  = \nonumber \\
&=& {\rm e} v_F \hat{\Psi}^\dagger(x) \left[ \sigma_z \cos\theta(x)  -\sigma_y \sin\theta(x) \right] \hat{\Psi}(x) \,\,\,.\,\,\,
\end{eqnarray}
Integrating the   Heisenberg equation for the field operator  
\begin{equation}\label{Heisenberg}
\partial_x \left(  e^{i \theta(x) \sigma_x/2} \hat{\Psi}(x)\right)= - \frac{\sigma_z}{v_F}    e^{i \theta(x) \sigma_x/2} \partial_t \hat{\Psi}(x) 
\end{equation}
around   any point   $x$, including  possible discontinuity points of $\theta(x)$, the following boundary condition is found
\begin{equation}\label{bc-model-2}
e^{i \theta(x^+) \sigma_x/2} \hat{\Psi}(x^+) =e^{i \theta(x^-) \sigma_x/2} \hat{\Psi}(x^-) \quad,
\end{equation}
which in turn straightforwardly implies the continuity of the current operator (\ref{curr-model-2}). In particular, for a step-like model $\theta(x<0)=0$ and $\theta(x>0)=\pi$ of an interface located at $x_0=0$,  Eq.(\ref{bc-model-2}) reduces to
\begin{equation}\label{bc-step-model-2}
\begin{array}{lcl} 
\hat{\Psi}_{\uparrow}(0^+) &=& i \hat{\Psi}_{\downarrow}(0^-) \\
\hat{\Psi}_{\downarrow}(0^+) &=& i \hat{\Psi}_{\uparrow}(0^-)
\end{array}
\end{equation}
and describes a spin-rotation process occurring at the interface.
Differently from model 1, the Heisenberg equation (\ref{Heisenberg})  does admit scattering state solutions $\hat{\Psi}_{E,\pm}(x,t)=\exp[-i E t/\hbar] \psi_{E\pm}(x)  \hat{a}_{E,\pm}/\sqrt{2\pi \hbar v_F}$, where $ \hat{a}_{E,\pm}$ are the energy-$E$ mode operators for scattering from the left(+) and from right(-), respectively, and $\psi_{E\pm}(x)$ are the related scattering wavefunctions.   For instance, the scattering state from left is given by
\begin{equation}
\begin{array}{lcl}
\psi_{E +}(x<0)&=& \left(\begin{array}{c} 1 \\ 0 \end{array} \right) e^{i E x/\hbar v_F} + r \left(\begin{array}{c} 0 \\ 1 \end{array} \right) e^{-i E x/\hbar v_F} \\
\psi_{E +}(x>0)&=& t \left(\begin{array}{c} 0 \\ 1 \end{array} \right) e^{i E x/\hbar v_F}  
\end{array}
\end{equation}
and, when inserted in Eq.(\ref{bc-step-model-2}), straightforwardly implies $r=0$ and $t=i$, leading to a perfect transmission $T=|t|=1$. 

In summary,   model 2 trivially solves the Dirac paradox  by simply introducing   spin-rotation processes at the interface.

\section{Dirac Heterojunctions with massless and massive modes}
\label{sec-3}
So far, we have considered  heterojunctions that purely involve Dirac massless modes and we have obtained two opposite answers to the Dirac paradox, depending on how the helicity change across the interface is modelled. Model 1, based on an inhomogeneous scalar velocity profile, implies that physical solutions necessarily involve injections from both sides of the junction and predicts no current flowing through the system, whereas model~2 ``circumvents" the paradox by introducing a spin-active interface. 
In this section we propose a model that, {\it without}  introducing any direct  spin-rotation processes at the interface,   leads to a  non-vanishing transmission.  

\begin{figure}[h]
\centering
\includegraphics[width=0.9\linewidth]{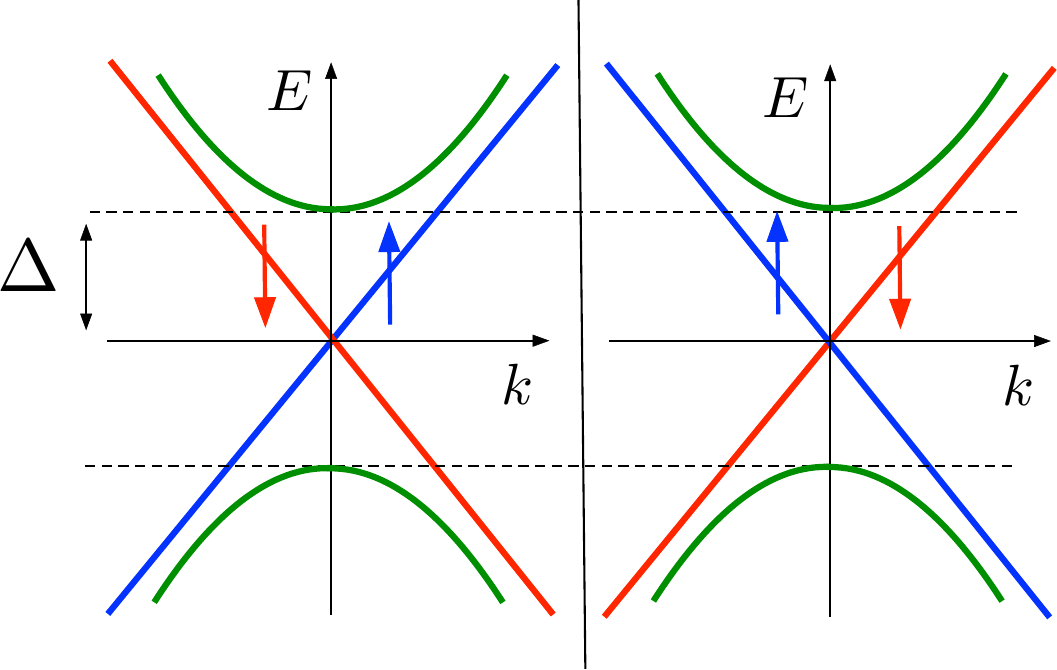}
\caption{\label{Fig2} The band spectrum of the massless+massive Dirac model with opposite helicity across an interface. While blue and red curves denote the spectrum of the massless modes, as in Fig.\ref{Fig1}, the green curves describe the spectrum of the massive modes, characterized by a gap~$2\Delta$. }
\end{figure}

Suppose that, along with the massless propagating Dirac fermions   illustrated in Fig.\ref{Fig1}, the system is also characterized by massive Dirac fermions, as sketched by the green curves of Fig.\ref{Fig2}.  Specifically, the model we  consider is
\begin{eqnarray}
\hat{\mathcal{H}}  &=& v_F \int \hat{\Psi}^\dagger(x) U^\dagger(x) \tau_z \sigma_z p_x \left(U^{}(x)\, \hat{\Psi}^{}(x)\right)\, dx + \nonumber \\
& & - \frac{\Delta}{2}  \int \hat{\Psi}^\dagger(x)   (\tau_0-\tau_z) \sigma_x  \, \hat{\Psi}^{}(x)\, dx  \label{Ham-model-3}
\end{eqnarray}
where $\hat{\Psi}=(\hat{\xi}_\uparrow, \hat{\xi}_\downarrow,\hat{\eta}_\uparrow,\hat{\eta}_\downarrow)^T$, with $\hat{\xi}_\uparrow, \hat{\xi}_\downarrow$ and $\hat{\eta}_\uparrow,\hat{\eta}_\downarrow$ denoting the massless  and massive fields, respectively. Here $\sigma_0$ and $\boldsymbol{\sigma}=(\sigma_x,\sigma_y,\sigma_z)$ denote the $2\times2$ identity matrix and Pauli matrices acting on the spin space, whereas $\tau_0$ and $\boldsymbol{\tau}=(\tau_x,\tau_y,\tau_z)$ the corresponding quantities acting on the massless-massive degree of freedom, which we shall label as pseudospin. In the first term of Eq.(\ref{Ham-model-3}) the $4 \times 4$ matrix $U(x)$   interpolates from $U_L$ on the left of the interface to its value $U_R$ on the right, where $U_{L/R}$ are  required to fulfill the following properties
\begin{eqnarray} 
U^\dagger_L \tau_z \sigma_z U^{}_L &=&+\tau_z \sigma_z \label{cond1-UL} \\
U^\dagger_R \tau_z \sigma_z U^{}_R &=&- \tau_z \sigma_z \quad, \label{cond1-UR}
\end{eqnarray}
so that   the $\hat{\xi}_\uparrow, \hat{\xi}_\downarrow$ modes have helicity  $+1$  on the left   of the interface and $-1$ on the right,  just like in the Dirac paradox configuration of Fig.\ref{Fig1}, whereas the opposite occurs for the   $\hat{\eta}_\uparrow,\hat{\eta}_\downarrow$ modes. 
The simplest example of a $U(x)$-matrix  fulfilling the conditions (\ref{cond1-UL})-(\ref{cond1-UR})  is $U(x)=\exp[i \theta(x) \tau_x \sigma_0 /2]$, where   $\theta(x)$ is a space-dependent angle describing a  rotation in {\it pseudospin} space around $\tau_x$ from  $\theta_L=0$ to $\theta_R=\pi$ and causing the  helicity flip, just like the spin-active model (\ref{Ham-model-2}) introduces a rotation in spin space. As we shall see below, there exists in fact a much broader set of possible choices for $U(x)$ that turn out to describe interesting and realistic cases.
The second term in Eq.(\ref{Ham-model-3}) describes the mass term for $\hat{\eta}_\uparrow$ and $\hat{\eta}_\downarrow$, and we shall be interested in the energy window $|E|< \Delta$ inside their gap, where these massive modes carry no current. 

In view of Eqs.(\ref{cond1-UL})-(\ref{cond1-UR}), the current operator related to the Hamiltonian (\ref{Ham-model-3})  
\begin{eqnarray}\label{curr-model-3}
\hat{J}(x)&=& {\rm e} v_F \hat{\Psi}^\dagger(x) U^\dagger(x) \tau_z   \sigma_z  \,  U(x) \hat{\Psi}(x)   
\end{eqnarray}
takes opposite expressions $\hat{J}_{L/R} =\pm  {\rm e} v_F \hat{\Psi}^\dagger(x) \tau_z   \sigma_z  \,  \hat{\Psi}(x) $ at the two sides of the interface. However, the boundary condition  
\begin{equation}
U(x^+) \hat{\Psi}(x^+) =U(x^-) \hat{\Psi}(x^-) \quad,
\end{equation}
obtained from integration of the  Heisenberg equation around any point $x$, guarantees that the current is in fact continuous for any $U(x)$. In particular, adopting again a step-like model $U(x)=U_L\, {\rm H}(-x)+ U_R\, {\rm H}(x)$ for an interface located at $x_0=0$, with ${\rm H}(x)$ denoting the Heaviside function, the field $\hat{\Psi}$  fulfills the interface boundary condition
\begin{equation}\label{Psi-M-Psi}
\hat{\Psi}(0^+) =\mathsf{M} \,\hat{\Psi}(0^-)  \quad,
\end{equation}
where 
\begin{equation}\label{M-def}
\mathsf{M}=U^{-1}_R U^{}_L
\end{equation} 
is the transfer matrix, which must  fulfill
\begin{eqnarray} \label{cond1-M}
 \mathsf{M}^\dagger \tau_z \sigma_z \mathsf{M} &=&- \tau_z \sigma_z \hspace{1cm} \mbox{\small ({\it Requirement \#1})}
\end{eqnarray}
as a straightforward  consequence of Eqs.(\ref{cond1-UL})-(\ref{cond1-UR}).  
Note that  Eq.(\ref{Psi-M-Psi}) implies that the field $\hat{\Psi}$  is discontinuous, 
as is customary for Dirac models in the presence of a $\delta(x)$-term, which in this case originates from $p_x U(x)$ term in the Hamiltonian  (\ref{Ham-model-3}).

Importantly,  in order to avoid trivial solutions to the Dirac paradox like in model (\ref{Ham-model-2}), we require that the model (\ref{Ham-model-3}) does not  directly introduce any spin-rotation process at the interface. This leads to impose   another requirement on the transfer matrix Eq.(\ref{Psi-M-Psi}), namely that~$\mathsf{M}$ is diagonal in spin space, i.e. 
\begin{equation}\label{cond2-M}
\begin{array}{l}
\mathsf{M}\,\, \, \mbox{must involve} \\
\mbox{only $\sigma_0$ and $\sigma_z$} 
\end{array} \hspace{1cm} \mbox{\small ({\it Requirement \#2})} \quad.
\end{equation}
It can be shown (see Appendix~\ref{AppA} for details) that the most general matrix fulfilling the requirements Eqs.(\ref{cond1-M})-(\ref{cond2-M}) has the following form in the $\tau \otimes \sigma$ basis
\begin{widetext}
\begin{eqnarray}\label{M-ris}
\mathsf{M}=    \left( \begin{array}{cccc}
i \beta_\uparrow \, e^{i (\nu_\uparrow-\gamma_\uparrow)} & 0 &  (1-i \beta_\uparrow) e^{i (\nu_\uparrow+\chi_\uparrow)} & 0\\
0 &  i \beta_\downarrow e^{i (\nu_\downarrow-\gamma_\downarrow)} & 0 &  (1-i \beta_\downarrow)e^{i (\nu_\downarrow+\chi_\downarrow)}\\
(1+i \beta_\uparrow) e^{i(\nu_\uparrow-\chi_\uparrow)}  &0  & -i \beta_\uparrow e^{i (\nu_\uparrow+\gamma_\uparrow)} & 0    \\
0 &(1+i \beta_\downarrow)  e^{i (\nu_\downarrow-\chi_\downarrow)}  &0  &  -i \beta_\downarrow e^{i(\nu_\downarrow+\gamma_\downarrow)}      
\end{array}
\right) 
\end{eqnarray}
\end{widetext}
and depends on 8 parameters, namely 4 real parameters   $\chi_\sigma,\gamma_\sigma,\beta_\sigma,\nu_\sigma$ for each spin sector $\sigma=\uparrow,\downarrow$. The vanishing entries in Eq.(\ref{M-ris}) encode the decoupling of the two spin sectors dictated by Eq.(\ref{cond2-M}).

\subsection{Scattering states}
Let us now focus on $E=0$, i.e. on  the middle of the massive energy gap, and build up scattering state  solutions on both sides of the junction, namely
{\small
\begin{eqnarray}
\lefteqn{ \hat{\Psi}(x<0) =} &&  \label{PsiL}   \\
&=&  \hat{a}_{L\uparrow} \left( \begin{array}{c} 1 \\0 \\ 0 \\0 
\end{array}
\right) e^{i k_0 x} \, + \hat{b}_{L\downarrow} \left( \begin{array}{c} 0 \\1  \\0 \\0 
\end{array}
\right)  e^{-i k_0 x}+ \frac{\hat{c}_L}{\sqrt{2}}  \left( \begin{array}{c} 0 \\0  \\-i \\1 
\end{array}
\right) e^{\kappa_0 x} \nonumber 
\end{eqnarray}
}
and
{\small
\begin{eqnarray}
\lefteqn{ \hat{\Psi}(x>0) =} &&  \label{PsiR} \\
&=& \hat{a}_{R\uparrow} \left( \begin{array}{c} 1 \\0 \\ 0 \\0 
\end{array}
\right)   e^{-i k_0 x} + \hat{b}_{R\downarrow} \left( \begin{array}{c} 0 \\1 \\ 0 \\0 
\end{array}
\right)   e^{i k_0 x}    + \frac{\hat{c}_R}{\sqrt{2}}  \left( \begin{array}{c} 0 \\0  \\-i \\1 
\end{array}
\right) e^{-\kappa_0 x} \nonumber
\end{eqnarray}
}

\noindent where $k_0=0$, $\kappa_0=\Delta/\hbar v_F$. Here $\hat{a}_{L\uparrow}$ and $\hat{a}_{R\uparrow}$ are incoming operators describing a propagating mode impinging from the left(L) and from the right(R) of the interface, respectively, whereas $\hat{b}_{L\downarrow}$ and $\hat{b}_{R\downarrow}$ are outgoing operators for  modes propagating to the left and to the right, respectively. Note that in the Dirac paradox configuration (see Fig.\ref{Fig2}) incoming states and outgoing states have opposite spin, namely spin-$\uparrow$ and  spin-$\downarrow$, respectively. Furthermore in Eqs.(\ref{PsiL}) and (\ref{PsiR})   $\hat{c}_{L}$ and $\hat{c}_{R}$ describe evanescent modes on the left- and on the right-hand side of the interface. Importantly, because they are massive, their spinors have two non vanishing components and their spin points along $y$. Introducing Eqs.(\ref{PsiL}) and (\ref{PsiR}) into  Eq.(\ref{Psi-M-Psi}) and using Eq.(\ref{M-ris}), one can write
\begin{equation}\label{out-in}
\left( \begin{array}{c}
\hat{b}_{L\downarrow} \\
\hat{b}_{R\downarrow} \\
\hat{c}_{L}\\
\hat{c}_{R}
\end{array} \right) =\begin{pmatrix}
&\mathsf{S} &\\
&\tilde{\mathsf{S}} &
\end{pmatrix} \left( \begin{array}{c} \hat{a}_{L\uparrow} \\ \hat{a}_{R\uparrow} \end{array} \right)\quad,
\end{equation}
where $\mathsf{S}$ denotes the Scattering Matrix returning the outgoing propagating modes
{\small
\begin{eqnarray}
\lefteqn{\mathsf{S}= \frac{i e^{-i   \Delta \chi }}{(1-i \beta_\uparrow) (1+i \beta_\downarrow)} \times } & &  \label{Smat} \\ & & \nonumber \\
& & \begin{pmatrix}
e^{i   \Delta \nu }+e^{-i   \Delta \gamma } \beta_\uparrow \beta_\downarrow &  i (e^{i (\gamma_\downarrow-\nu_\uparrow )} \beta_\downarrow -e^{i (\gamma_\uparrow -\nu_\downarrow)} \beta_\uparrow)\\ & \\
i (e^{i (\nu_\uparrow -\gamma_\downarrow)} \beta_\downarrow- e^{i (\nu_\downarrow-\gamma_\uparrow)} \beta_\uparrow) &   e^{-i   \Delta \nu }+e^{i   \Delta \gamma } \beta_\uparrow \beta_\downarrow 
\end{pmatrix} \nonumber
\end{eqnarray}
}

\noindent with $\Delta \chi \doteq \chi_\uparrow-\chi_\downarrow$, $\Delta \nu \doteq \nu_\uparrow-\nu_\downarrow$ and $\Delta \gamma \doteq \gamma_\uparrow-\gamma_\downarrow$, whereas
\begin{eqnarray}
\tilde{\mathsf{S}}= \frac{\sqrt{2} \,e^{-i  \chi_\uparrow}}{1-i\beta_\uparrow}    \begin{pmatrix}
 \beta_\uparrow e^{-i \gamma_\uparrow } & i e^{-i  \nu_\uparrow}  \\ & \\
i e^{+i \nu_\uparrow} &   \beta_\uparrow e^{i  \gamma_\uparrow }
\end{pmatrix} \label{Stildemat}
\end{eqnarray}
is the matrix yielding the evanescent modes.\\

In Eq.(\ref{out-in}), setting $\hat{a}_{R\uparrow}\rightarrow 0$ yields a scattering state with injection from left, while  a scattering state with injection from right is obtained for $\hat{a}_{L\uparrow}\rightarrow 0$.  Thus, differently from  model~1 in Eq.(\ref{Ham-model-1}),     the model in Eq.(\ref{Ham-model-3})  does allow for scattering solutions. The transmission coefficient   $T_0=|t_0|^2$, obtained from the off-diagonal entries of the Scattering Matrix (\ref{Smat}),  reads
\begin{equation}\label{T0-ris}
T_0=\frac{\beta_\uparrow^2+\beta_\downarrow^2-2 \beta_\uparrow \beta_\downarrow \cos\varphi}{(1+\beta_\uparrow^2)(1+\beta_\downarrow^2)}\quad,
\end{equation}
and depends on the three parameters $\beta_\uparrow$, $\beta_\downarrow$ and $\varphi=\Delta \gamma+\Delta\nu$.    
 To understand how the transmission between two propagating electronic states with oppositely oriented  spins is possible, let us  for instance set $\hat{a}_{R\uparrow} \rightarrow 0$ in Eq.(\ref{out-in}), which corresponds to a  scattering process where a   spin-$\uparrow$ state incoming   from the far left is transmitted into a spin-$\downarrow$ state outgoing to the far right. By inspecting the spin spatial profile of Eqs.(\ref{PsiL})-(\ref{PsiR}), one observes that  far away from the interface the total spin is mainly carried by the massless propagating states and is directed along the $z$-axis. However, near the interface, spin acquires also a component along $y$  because of the presence of the massive states   (third terms of Eqs.(\ref{PsiL})-(\ref{PsiR})).  Indeed  the  conservation of $S^{tot}_z= \hbar \,\hat{\Psi}^\dagger \tau_0 \sigma_z \hat{\Psi}^{}/2$ is broken precisely by the mass in the Hamiltonian Eq.(\ref{Ham-model-3}). Thus, when approaching the interface, the total spin   {\it rotates} in the $y$-$z$ plane, thereby  allowing the transmission from a   spin-$\uparrow$    to a   spin-$\downarrow$  massless state.
Note the essential difference with respect to model~2: There, the spin-rotation is induced {\it directly} on the massless modes by a spin-active interface [see Eq.(\ref{Ham-model-2})], whereas here  the transfer matrix in Eqs.(\ref{Psi-M-Psi}) and (\ref{M-ris}) is fully diagonal in spin  [see Eq.(\ref{cond2-M})] and the spin rotation occurs {\it indirectly}, i.e. through the coupling between massless and massive modes localized at the interface.\\

\begin{figure}[h]
\centering
\includegraphics[width=0.95\linewidth]{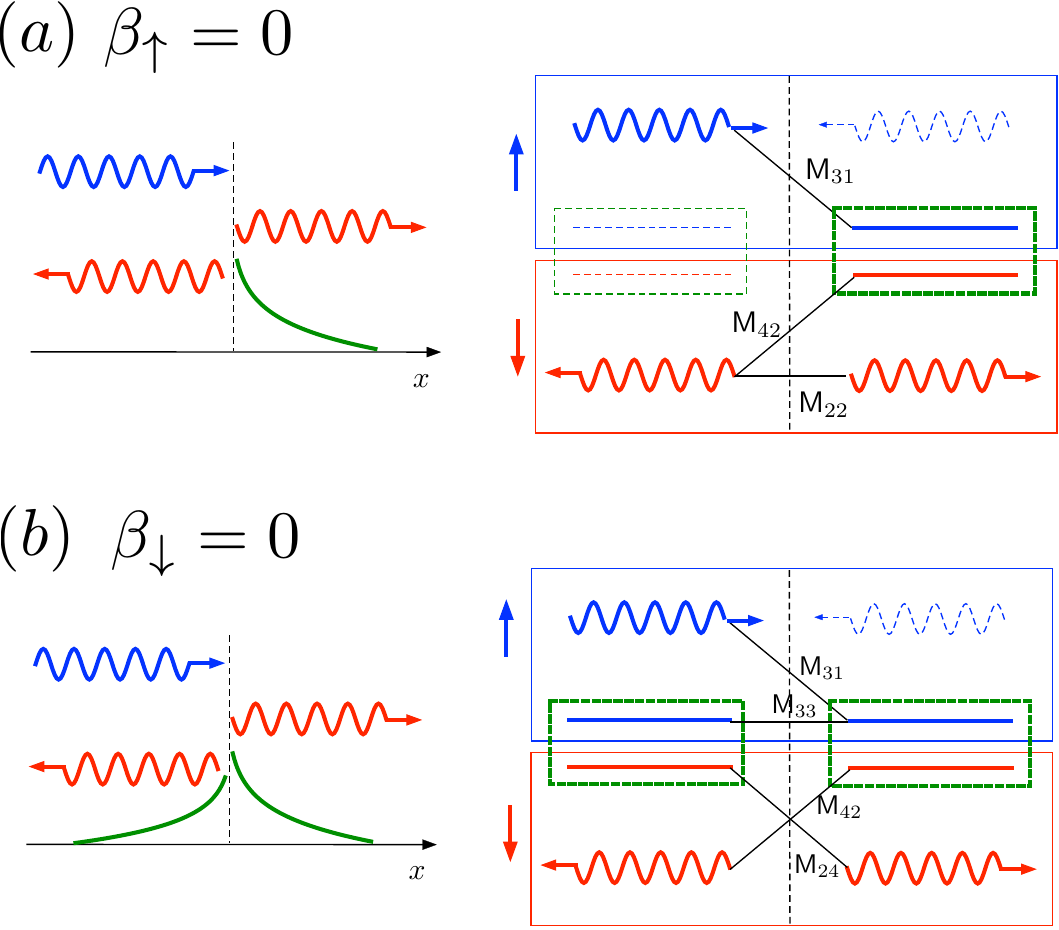}
\caption{\label{Fig3} 
For each panel, the left-hand side sketches the scattering state wavefunction in the case of injection from the left, resulting from Eqs.(\ref{out-in})-(\ref{Smat})-(\ref{Stildemat}) for $\hat{a}_{R\uparrow}\rightarrow 0$. Blue and red wiggy lines describe spin-$\uparrow$ and spin-$\downarrow$ propagating massless states, respectively, whereas solid green lines describe the evanescent wave of the massive mode. The right-hand side of each panel is a graphical representation of Eq.(\ref{Psi-M-Psi}), where black lines represent the transfer matrix entries connecting the non vanishing components of such scattering state.    (a) The case with $\beta_\uparrow=0$. The evanescent mode is present only on the right side of the interface. Here $\mathsf{M}_{31}=1$, $M_{42}=1+i\beta_\downarrow$ and $\mathsf{M}_{22}=i\beta_\downarrow$. (b) The case with $\beta_\downarrow=0$.  In this case the evanescent modes are present on both sides of the junction. Here $\mathsf{M}_{31}=1+i\beta_\uparrow$, $M_{33}=-i\beta_\uparrow$ and $\mathsf{M}_{24}=\mathsf{M}_{42}=1$. In all cases,  despite the transfer matrix only connects states with the same spin, the presence of the evanescent modes of the massive field enables  a spin-flip transmission between the propagating modes. }
\end{figure}

To a more formal level, the process can be illustrated in terms  of the Transfer Matrix as follows. Let us again  consider for definiteness the   scattering   from left, i.e. $\hat{a}_{R\uparrow} \rightarrow 0$ in Eq.(\ref{out-in}), and also set for simplicity   all phases    to zero ($\gamma_\sigma=\chi_\sigma=\nu_\sigma=0$)  in Eqs.(\ref{M-ris}),  (\ref{Smat}) and (\ref{Stildemat}). We first focus on  the case $\beta_\uparrow=0$, where the scattering state resulting from   Eqs.(\ref{out-in})-(\ref{Smat})-(\ref{Stildemat}) is  sketched on the left-hand side of Fig.\ref{Fig3}(a): The blue (red) wiggy line describes the incoming spin-$\uparrow$ state (outgoing spin-$\downarrow$ states), while an evanescent wave (green solid line) is present only for $x>0$. Its role is elucidated on the  right hand side of Fig.\ref{Fig3}(a), which is a graphical representation of Eq.(\ref{Psi-M-Psi}) where the non-vanishing components of such a scattering state are connected  across the interface by the transfer matrix entries (black lines).
When the massless spin-$\uparrow$ state propagates towards the interface from the left,  the transfer matrix Eq.(\ref{M-ris}) connects it  through the entry  $\mathsf{M}_{31}=1$ to its massive evanescent  partner with the same spin located across the interface, represented by a green dashed box, with the thick solid lines inside it denoting its two spin components. Because such a mode is massive,   inside the gap it always exhibits {\it both} spin components [see third term  in Eq.(\ref{PsiR})]. Thus,  its spin-$\downarrow$ component is also present and is connected through the transfer matrix entry $\mathsf{M}_{42}=1+i\beta_\downarrow$ to its spin-$\downarrow$ massless partner, which   describes the reflected wave propagating to the left of the junction. Finally, the latter is also coupled, through the entry $\mathsf{M}_{22}=i\beta_\downarrow$, to the massless spin-$\downarrow$ state outgoing to the right of the junction. Thus, despite the interface connects   only   states with the same spin on the two sides,   the presence of an evanescent massive mode exhibiting both spin components leads to an effective spin-flip transmission between massless modes.  

Let us now consider the case $\beta_\downarrow=0$. In this case the scattering state resulting from the  solution  Eqs.(\ref{out-in})-(\ref{Smat})-(\ref{Stildemat}) exhibits evanescent modes on both sides of the junction, as sketched in the left hand side of Fig.\ref{Fig3}(b). The scheme on the right hand side of Fig.\ref{Fig3}(b) illustrates the related Eq.(\ref{Psi-M-Psi}). While the entry $\mathsf{M}_{31}$ is modified to $\mathsf{M}_{31}=1+i\beta_\uparrow$,   a connection $\mathsf{M}_{33}=-i\beta_\uparrow$ opens up across the junction between the two spin-$\uparrow$ components of the massive modes. In turn, their corresponding spin-$\downarrow$ components are connected through the entries $\mathsf{M}_{24}=\mathsf{M}_{42}=1$ to the spin-$\downarrow$ massless modes across the junction, thereby inducing again spin-flipped reflection and transmission.

The general case, where both $\beta_\uparrow$ and $\beta_\downarrow$ are non vanishing, is a combination of the  two elementary cases and yields the transmission coefficient (\ref{T0-ris}).  Note that in the limit where both $\beta_\uparrow\rightarrow 0$ and $\beta_\downarrow\rightarrow 0$, the transmission coefficient (\ref{T0-ris}) vanishes. This can also be understood by realizing that in such limit  the transfer matrix (\ref{M-ris}) reduces to $\mathsf{M} = \tau_x \sigma_0$, yielding the boundary conditions
 \begin{equation}\label{bc-step-model-3}
\left\{ \begin{array}{lcl} 
\hat{\xi}_\sigma (0^+) &=&   \hat{\eta}_\sigma(0^-) \\  
 \hat{\eta}_\sigma(0^+) &=& \hat{\xi}_\sigma (0^-) 
\end{array} \right. \hspace{1cm} \sigma=\uparrow,\downarrow \quad,
\end{equation}
so that e.g. a massless mode incoming from the left towards the interface is completely transformed   into its massive evanescent mode partner across the interface (with the same spin), which carries no current.\\

In summary, although massive modes do not carry any current inside the gap, their presence is important in inhomogeneous problems because they may localize at the interfaces. In particular they are crucial  in  the Dirac paradox, for they provide an indirect coupling between the two spin channels   that would be otherwise uncoupled by the interface transfer matrix. This leads to an effective spin-flip transmission of the massless propagating modes. Moreover, in contrast with the  models 1 and 3 discussed in Sec.\ref{sec-2}, here the transmission coefficient is tunable from 0 to 1 through the 3 knobs $\beta_\uparrow, \beta_\downarrow$ and $\varphi$.
This is one of the main results of our paper.

\section{Spin-orbit coupled nanowires}
\label{sec-4}
In this section we shall show that  the model  presented in Sec.\ref{sec-3} can be realized with spin-orbit coupled NWs, under suitable circumstances. First, we shall briefly recall how  these systems, when exposed to an external magnetic field, can host helical states  described by Dirac massless fermions, as well as gapped Dirac states. Then, focussing on energies  inside the gap opened up by the magnetic field, we shall explicitly derive the effective low-energy model for these systems. Finally, we shall consider   an inhomogeneous spin-orbit coupling profile that, in suitable regimes,   realizes the   Dirac paradox configuration involving both massless and massive modes, just like in  the model proposed above.

\subsection{The NW Hamiltonian and its low energy limit}
\label{sec-4A}
We consider a ballistic single-channel semiconductor  NW  deposited on a substrate. For   NWs like InSb or InAs, the structural inversion asymmetry  can lead to   quite strong  RSOC\cite{nilsson_2009,xu_2012,kouwenhoven_PRL_2012,ensslin_2010,gao_2012,joyce_2013}, which can further be  tuned with appropriate gating techniques. Furthermore we assume that a uniform magnetic field is applied parallel to the nanowire axis,  denoted by $x$, while   the substrate plane will be identified as $x$-$z$.\\

We shall adopt a widely used model to describe the NW\cite{streda,vonoppen_2010,dassarma_2010,loss_PRB_2011,lutchyn_2012,loss_PRB_2017}, whose main ingredients are summarized here below, while details are reported in Appendix \ref{AppB} for the sake of completeness. The  second-quantized NW Hamiltonian    consists of three terms $\hat{\mathcal{H}}_{NW}=\hat{\mathcal{H}}_{kin}+\hat{\mathcal{H}}_{R}+\hat{\mathcal{H}}_{Z}$ and can be written as
$
\hat{\mathcal{H}}_{NW} = \int \hat{\Phi}^\dagger(x)\,H_{NW}(x)\,  \hat{\Phi}(x)\,dx
$. Here $\hat{\Phi}(x)= ( \hat{\Phi}_\uparrow(x) \,,\,\hat{\Phi}_\downarrow(x)  )^T$ is the electron spinor field, with $\uparrow,\downarrow$ corresponding to spin projections along~$z$, and
\begin{equation} \label{H(x)_homo}
H_{NW}(x)=  \frac{p_x^2}{2 m^*} \sigma_0 -\frac{\alpha}{\hbar}p_x  \sigma_z\, \, - h_x \sigma_x   
\end{equation}
contains the kinetic term characterized by an effective mass $m^*$, the Rashba term with a  RSOC  $\alpha$, and the Zeeman term describing the coupling   $h_x=g \mu_B B_x/2$   with the external magnetic field $\mathbf{B}=(B_x,0,0)$, with $\mu_B$ denoting the Bohr magneton and $g$ the NW Land\'e  factor. 
The model is characterized by two energy scales, namely the spin-orbit energy   
\begin{equation}\label{ESO-def}
E_{SO}=\frac{m^* \alpha^2}{2 \hbar^2} \quad,
\end{equation}
and the Zeeman energy 
\begin{equation}
E_Z=|h_x| \label{EZ-def} \quad.
\end{equation} 
For definiteness, we shall henceforth assume $h_x>0$ and identify $h_x=E_Z$. 
The spin-orbit wavevector
\begin{equation}\label{kSO-def}
k_{SO} =\frac{\sqrt{{2m^*E_{SO}}}}{ \hbar} = \frac{|\alpha| m^*}{\hbar^2} \quad,
\end{equation} 
and the  Zeeman wavevector 
\begin{equation}\label{kZ-def}
k_{Z} =\sqrt{{2m^*E_{Z}}}\, / \hbar  
\end{equation} 
are the wavevectors associated to such energies. Diagonalizing the model in momentum space, one obtains a spectrum characterized by two energy bands separated at $k=0$ by a  gap $2E_Z$ centered around the  energy $E=0$.
Depending on the ratio of $E_Z$ to $E_{SO}$, the  qualitative behavior of these bands is different. In fact, two regimes can be identified: 
(a) for $E_Z > 2E_{SO}$ (Zeeman-dominated regime)  both bands have a minimum at $k=0$ taking values $E_{\pm}^{\rm min}=\pm E_Z$, whereas 
(b) for $E_Z < 2E_{SO}$ (Rashba-dominated regime)  the upper band still has a minimum $E_{+}^{\rm min}=+E_Z$ at $k=0$, while the lower band has a local maximum at $k=0$ and acquires two degenerate minima $E_{-}^{\rm min}=-E_{SO}-E_Z^2/4 E_{SO}$ at $k = \pm k_{SO} \sqrt{1-E_Z^2/4 E_{SO}^2}$. 

\begin{figure}[h]
\centering
\includegraphics[width=0.9\linewidth]{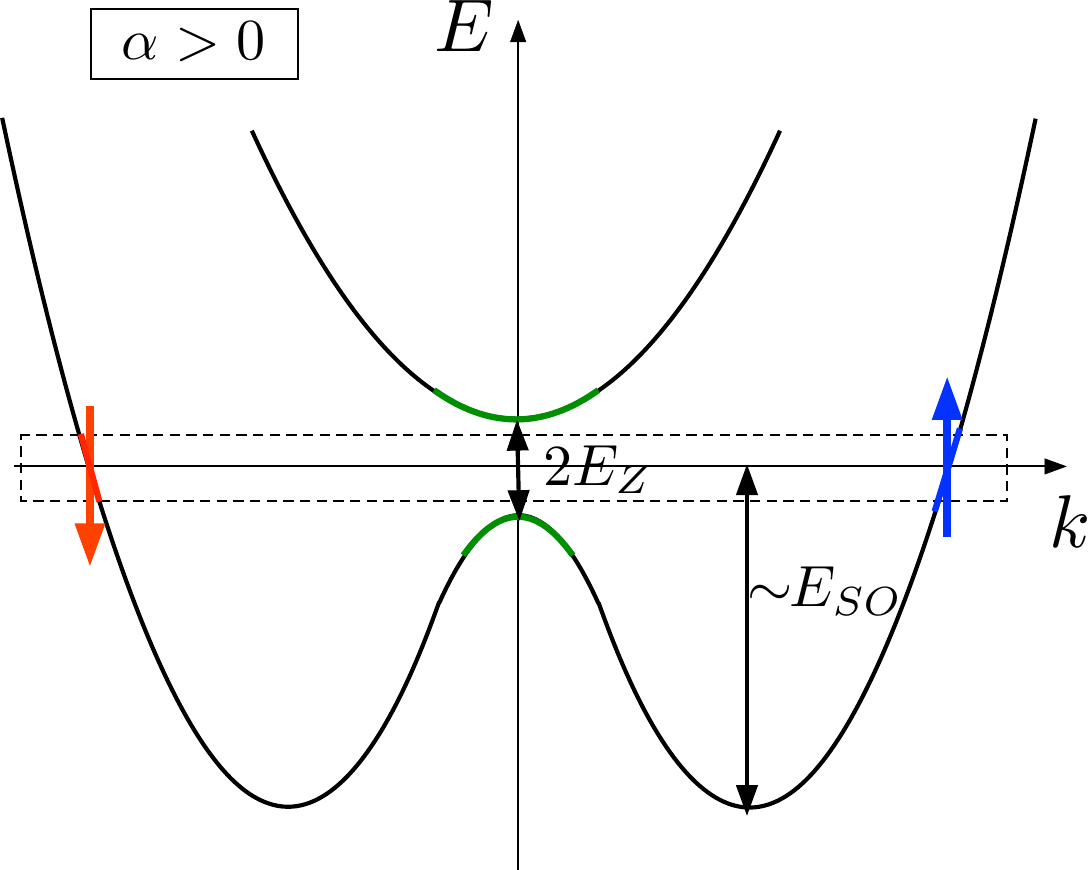}
\caption{\label{Fig4} In the deep Rashba-dominated regime ($2E_{SO} \gg E_Z$) and for energies $|E| \ll E_Z$,  a NW exposed to an external magnetic field exhibits   helical states near the Fermi points $\pm 2 k_{SO}$. Their spin orientation is locked to the propagation direction and is determined by the sign of the RSOC $\alpha$. The spin orientation  is shown for the case $\alpha>0$. The low energy massless and massive modes are highlighted with the same colors as in Fig.\ref{Fig4}.}
\end{figure}

In   the following we shall focus on the  deep Rashba-dominated regime ($E_Z \ll 2 E_{SO}$), illustrated in   Fig.\ref{Fig4}, and analyze  the energy window inside the magnetic gap ($|E|\ll E_Z$), highlighted by the dashed box. As is well known,
  in this range  the  NW   propagating eigenstates are helical\cite{streda,vonoppen_2010,dassarma_2010,loss_PRB_2011,lutchyn_2012,loss_PRB_2017,depicciotto_2010,kouwenhoven_nanolett_2013,kouwenhoven_natcom_2017}: Their   dispersion relation   is well described by a linear behavior  near the Fermi points $k\simeq \pm 2k_{SO}$, while their spin orientation, mainly dictated by the Rashba term,   is locked to the propagation direction. For  $\alpha>0$, right-moving electrons near the right Fermi point $k\simeq +2 k_{SO}$ are characterized by spin-$\uparrow$, while left-moving electrons near the left Fermi point $k\simeq -2 k_{SO}$ have spin-$\downarrow$   (see  Fig.\ref{Fig4}). The opposite occurs if $\alpha<0$. The dynamics of these low energy propagating modes, which we shall denote by $\hat{\xi}_\uparrow$, $\hat{\xi}_\downarrow$, is thus described by a massless Dirac Hamiltonian.  Note that the presence of one single Dirac cone is not an artifact of the continuum model (\ref{H(x)_homo})  and can be found also in a regularized lattice version of it (see Appendix \ref{AppC}).  Importantly, the helicity of the Dirac cone is determined by the \textit{sign}  of the RSOC $\alpha$
  \begin{equation}
\label{sign-alpha-def}
s_\alpha=\mbox{sgn}(\alpha) \quad.
\end{equation}
This suggests that a junction between two NW regions with opposite values of RSOC  realizes the Dirac paradox configuration.

However,  as highlighted by the green lines in Fig.\ref{Fig4}, the NW also exhibits low-energy gapped modes  near $k=0$, whose spin components shall be denoted as $\hat{\eta}_\uparrow$ and $\hat{\eta}_\downarrow$. Notably, these modes turn out to behave as  massive Dirac fermions with a mass term $\Delta=E_Z$.  In problems involving homogeneous NWs these modes are dropped because they are not normalizable. However, as observed in Sec.\ref{sec-3}, in inhomogeneous problems such as the Dirac paradox configuration they describe evanescent waves that,   despite   carrying no current, ensure the wavefunction matching at the interface. For these reasons, the effective low energy theory  capturing the physical properties of the Dirac paradox configuration realized with NWs is a Dirac model involving both massless and massive modes.\\

To derive such effective theory describing low energy excitations $|E|\ll E_Z \ll 2E_{SO}$, we assume that the ground state is the  Fermi sea where all NW states below the midgap energy ($E<0$) are occupied, and we perform an expansion near the points $k\simeq \pm 2k_{SO}$ and $k\simeq 0$. It is possible to show (details can be found in Appendix~\ref{AppB}), that the low energy excitations of the NW  Hamiltonian are equivalent to low energy excitations of the massless+massive Dirac model 
\begin{eqnarray}
   \hat{\mathcal{H}}_{NW}  =  \sum_{q = - \infty}^{+\infty} 
     (\hat{\xi}_{q  \uparrow}^\dagger\ \hat{\xi}_{q  \downarrow}^\dagger )
    \begin{pmatrix}
    \hbar s_\alpha v_{so} q && 0\\
    0 && - \hbar s_\alpha v_{SO} q
    \end{pmatrix}
     \left( \begin{array}{c} \hat{\xi}_{q  \uparrow}\\ \hat{\xi}_{q  \downarrow}\end{array} \right)   \nonumber \\
  \displaystyle + \sum_{q = - \infty}^{+\infty} 
  (\hat{\eta}_{q  \uparrow}^\dagger\ \hat{\eta}_{q  \downarrow}^\dagger)
    \begin{pmatrix}
    - \hbar s_\alpha v_{so} q && -E_Z\\
    -E_Z&& \hbar s_\alpha v_{SO} q
    \end{pmatrix}
    \Vector{\hat{\eta}_{q  \uparrow}\\ \hat{\eta}_{q  \downarrow}}   \hspace{1.cm} \label{low-energy-model}
\end{eqnarray}
where $v_{SO}=\hbar k_{SO}/m^*$. 
Introducing the low-energy fields ($\sigma =   \uparrow,  \downarrow $)
\begin{equation}
    \hat{\xi}_{\sigma}(x) = \frac{1}{\sqrt{\Omega}} \sum_q \hat{\xi}_{q, \sigma} e^{iqx} \hspace{0.3cm} , \hspace{0.3cm} \hat{\eta}_{\sigma}(x) = \frac{1}{\sqrt{\Omega}} \sum_q \hat{\eta}_{q, \sigma} e^{iqx}
\end{equation}
that physically describe excitations  varying over lengthscales much longer than the spin-orbit length $l_{SO}=k_{SO}^{-1}$, the NW Hamiltonian can be expressed as
\begin{eqnarray}
 \lefteqn{\hat{\mathcal{H}}_{NW}  =  } & &\label{HNW-lowene} \\
 &=& \int dx 
    \hat{\Psi}^\dagger(x)
    \left(   s_\alpha v_{SO}  \tau_z \sigma_z p_x - \frac{E_Z}{2}(\tau_0 - \tau_z) \sigma_x \right)
    \hat{\Psi}(x) \nonumber
\end{eqnarray}
where  $\hat{\Psi}(x)=(\hat{\xi}_\uparrow, \hat{\xi}_\downarrow,\hat{\eta}_\uparrow,\hat{\eta}_\downarrow)^T$ is a 4-component spinor field. One can now realize the connection between the NW Hamiltonian (\ref{HNW-lowene}) and the model  introduced in Sec.\ref{sec-3} in Eq.(\ref{Ham-model-3}). Indeed,     identifying   $v_{SO}\rightarrow v_F$ and $E_Z \rightarrow \Delta$, Eq.(\ref{HNW-lowene}) describes one side of the junction  model (\ref{Ham-model-3}), where the sign $s_\alpha$  of the RSOC [see Eq.(\ref{sign-alpha-def})] implements the condition Eq.(\ref{cond1-UL}) or (\ref{cond1-UR}) and determines which side of the junction is described.

Finally, the original field $\hat{\Phi}$ can be expressed  in terms of the Dirac slowly varying     modes  $(\hat{\xi}, \hat{\eta})$   and the fast oscillating plane waves related to the midgap Dirac points, as follows
\begin{equation} \label{eq:Total_field} 
 \left( \begin{array}{c}   \hat{\Phi}_\uparrow(x) \\  \hat{\Phi}_\downarrow(x) 
\end{array}\right)  
 =  \left( \begin{array}{c} e^{+2 i s_\alpha k_{SO} x}\,\hat{\xi}_{\uparrow}(x) +  \hat{\eta}_{\uparrow}(x)\\
                    e^{-2 i s_\alpha k_{SO} x}\, \hat{\xi}_{\downarrow}(x) +  \hat{\eta}_{\downarrow}(x)\end{array} \right)  \quad.
\end{equation}

\subsection{The case of inhomogeneous RSOC}
\label{sec-4C}
\begin{figure}[h!]
  \includegraphics[width=\linewidth]{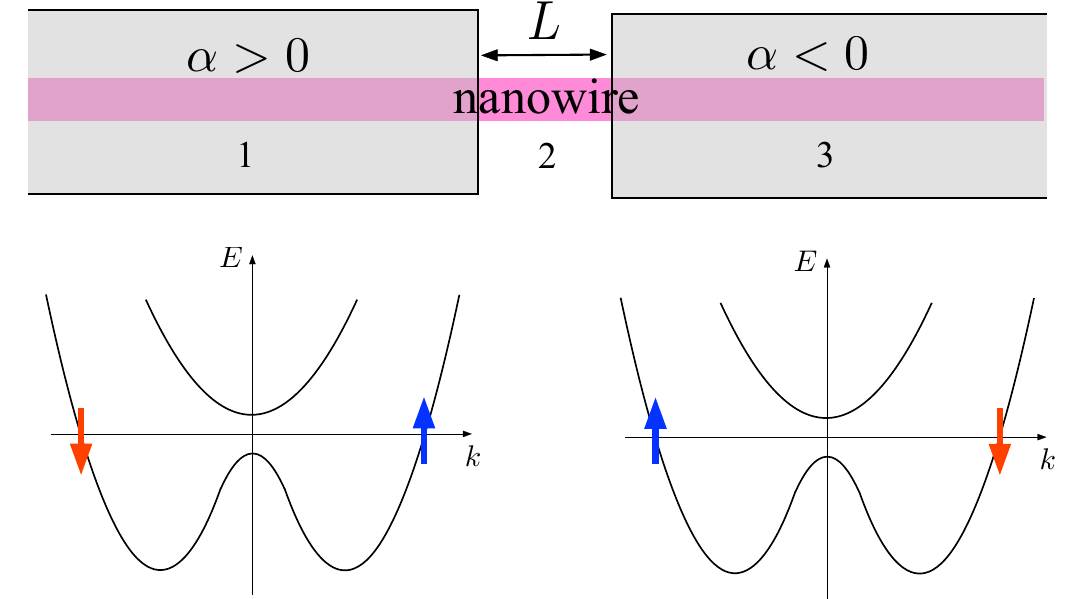}
  \caption{By gating different portions of the NW with metallic electrodes an inhomogeneous RSOC like (\ref{alpha-profile}) can  be realized. Inside the Zeeman gap, induced by an external   magnetic field applied along the nanowire axis, the helical states in the two outer regions have opposite helicity and thus realize the Dirac paradox configuration.}
  \label{Fig5}
\end{figure}

Because the helicity of the NW low energy massless modes is determined by the sign of the RSOC, one can envisage a setup where two different NW portions  are characterized by values of $\alpha$ with   opposite signs, as illustrated in Fig.\ref{Fig5}. Indeed 
the huge advances of gating techniques  enable one to realize different gate potentials to various portions of the nanowires\cite{sasaki_2013,micolich,sasaki_2017,das_2019,guo_2021,sasaki_2021}, thereby locally varying the magnitude  and even the sign of the RSOC\cite{gao_2012,slomski_NJP_2013,wimmer_2015,nygaard_2016,sherman_2016,tokatly_PRB_2017,loss_2018,goldoni_2018,tsai_2018,gao-review,lau_2021,kaindl_2005,wang-fu_2016,nitta-frustaglia}.
The overall system can thus be described by a \textit{inhomogeneous} spin-orbit coupling $\alpha(x)$ and the Hamiltonian (\ref{H(x)_homo}) is generalized to\cite{sanchez_2006,sanchez_2008,sherman_2011,sherman_2013,sherman_2017,loss_EPJB_2015,dolcini-rossi_PRB_2018,lorenzo_2020,rossi_EPJB}
\begin{equation} \label{H(x)_inhomo}
H(x)=  \frac{p_x^2}{2 m^*} \sigma_0 -\frac{\left\{ \alpha(x) , p_x\right\}}{2\hbar}  \sigma_z\, \, -  h_x \sigma_x 
\end{equation}
where the anticommutator is necessary since  $p_x$ does not commute with the space-dependent RSOC. In particular, as an elementary building block, one can consider a step-like RSOC profile $\alpha(x)=\alpha_L\, {\rm H}(x_0-x)+ \alpha_R\, {\rm H}(x-x_0)$ describing an interface located at $x_0$ between two regions with RSOC equal to $\alpha_L$  and $\alpha_R$. In such a configuration one can straightforwardly derive the following matching conditions\cite{lorenzo_2020}
\begin{equation}\label{bc-gen}
\begin{cases}
\hat{\Phi}_\uparrow (x_0^-) = \hat{\Phi}_\uparrow (x_0^+)\\
\hat{\Phi}_\downarrow (x_0^-) = \hat{\Phi}_\downarrow (x_0^+)\\
\partial_x \hat{\Phi}_\uparrow (x_0^-) = \partial_x \hat{\Phi}_\uparrow (x_0^+) -i\frac{m^*}{\hbar^2} ( \alpha_R-\alpha_L) \hat{\Phi}_\uparrow (x_0)\\
\partial_x \hat{\Phi}_\downarrow (x_0^-) = \partial_x \hat{\Phi}_\downarrow (x_0^+)+i \frac{m^*}{\hbar^2} ( \alpha_R-\alpha_L)\hat{\Phi}_\downarrow (x_0) \, .
\end{cases}
\end{equation}
This provides all the ingredients for a concrete implementation of the Dirac paradox. In order to be more realistic, we shall consider a three-region configuration where the RSOC varies as
\begin{equation}
\alpha(x)= \left\{ 
\begin{array}{lcll} 
+\alpha>0 & \mbox{for} & x<-L/2 & \mbox{\small (region 1)} \\
 0 & \mbox{for} & |x|<L/2 & \mbox{\small (region 2)} \\
-\alpha <0 & \mbox{for} & x>+L/2 & \mbox{\small (region 3)} 
\end{array}
\right. \quad, \label{alpha-profile}
\end{equation}
where the two outer regions 1 and 3 with opposite RSOC are both assumed in the deep Rashba-dominated regime ($2E_{SO}\gg E_Z$),  and are separated by the central crossover region 2 with length~$L$ and with vanishing RSOC, i.e. in the Zeeman-dominated regime (see Fig.\ref{Fig5}).

Applying the general interface condition (\ref{bc-gen}) to the two interfaces $x_1=-L/2$ and $x_2=+L/2$ of the piecewise constant profile  (\ref{alpha-profile}), one can match the NW wavefunction in the three regions   and obtain the solution for the NW scattering problem with standard techniques\cite{datta-book} (see Appendix~\ref{AppD}). Although the resulting transmission coefficient is numerically exact and available for arbitrary values of $E$, $E_{SO}$ and $E_Z$, it is not quite amenable. 
However,  in the  energy window $|E| \ll E_Z \ll 2E_{SO}$ where  the Dirac paradox emerges, an analytical expression   can be gained from the effective low energy model. To this purpose, one can insert the expression (\ref{eq:Total_field}) for the field $\hat{\Phi}$ in the outer Rashba-dominated regions into the  interface condition (\ref{bc-gen}) and obtain the   low energy boundary conditions at the left interface $x_1=-L/2$
\begin{equation}
\left\{
\begin{array}{l}
    e^{-i k_{SO} L}\hat{\xi}_{\uparrow}(x_1^-) +  \hat{\eta}_{\uparrow}(x_1^-) = \hat{\Phi}_{\uparrow}(x_1^+)\\
    e^{+ i k_{SO} L}\hat{\xi}_{\downarrow}(x_1^-) +  \hat{\eta}_{\downarrow}(x_1^-) = \hat{\Phi}_{\downarrow}(x_1^+)\\
   + i k_{SO} \left[ e^{- i k_{SO} L}\hat{\xi}_{\uparrow}(x_1^-) -  \hat{\eta}_{\uparrow}(x_1^-) \right] = \partial_x \hat{\Phi}_{\uparrow}(x_1^+)\\
    - i k_{SO} \left[ e^{ +i k_{SO} L}\hat{\xi}_{\downarrow}(x_1^-) -  \hat{\eta}_{\downarrow}(x_1^-) \right] = \partial_x \hat{\Phi}_{\downarrow}(x_1^+)
    \end{array}  \right. \label{bc-1bis} 
\end{equation}    
and at the right interface $x_2=+L/2$
\begin{equation}
\left\{
\begin{array}{l}
 \hat{\Phi}_{\uparrow}(x_2^-) = e^{-i k_{SO} L }\hat{\xi}_{\uparrow}(x_2^+) +  \hat{\eta}_{\uparrow}(x_2^+)\\
\hat{\Phi}_{\downarrow}(x_2^-) = e^{+ i  k_{SO} L}\hat{\xi}_{\downarrow}(x_2^+) +  \hat{\eta}_{\downarrow}(x_2^+)\\
    \partial_x \hat{\Phi}_{\uparrow}(x_2^-) = - i k_{SO} \left[ e^{- i k_{SO} L}\hat{\xi}_{\uparrow}(x_2^+) -  \hat{\eta}_{\uparrow}(x_2^+) \right]\\
    \partial_x \hat{\Phi}_{\downarrow}(x_2^-) = +i k_{SO} \left[ e^{+i k_{SO} L}\hat{\xi}_{\downarrow}(x_2^+) - \hat{\eta}_{\downarrow}(x_2^+) \right] \label{bc-2bis}
    \end{array}  \right. \quad, 
\end{equation}    
where, consistently with the low energy limit, we have neglected the derivatives $\partial_x\hat{\xi}$ and $\partial_x \hat{\eta}$ of the slowly varying fields  with respect to the term proportional to $k_{SO}$, since they are characterized by wavevectors $|q| \ll k_{SO}$.

In the central region 2, where only the Zeeman coupling is present, the field $\hat{\Phi}$   can be expressed as a linear combination of propagating and evanescent waves that are eigenfunctions of $\sigma_x$, so that for $|x|<L/2$
\begin{eqnarray}
    \hat{\Phi}(x) &= & \frac{\hat{h}_E}{\sqrt{2}} \Vector{1\\1} e^{i k_{2, E} x}+  \frac{\hat{g}_E}{\sqrt{2}}\Vector{1\\1}e^{-i k_{2, E} x}  \label{Phi2}\\
    &+& \frac{\hat{d}_E}{\sqrt{2}}\Vector{1\\-1}e^{\kappa_{2, E} x} +   \frac{\hat{f}_E}{\sqrt{2}}\Vector{1\\-1} e^{-\kappa_{2, E} x}\nonumber
\end{eqnarray}
 where $\hat{h}_E, \hat{g}_E, \hat{d}_E$ and $\hat{f}_E$ are mode operators, whereas $k_{2,E} = k_Z\sqrt{1+E/E_Z}$, $\kappa_{2,E} = k_Z\sqrt{1 -  E/E_Z}$ and $k_Z$ is given in Eq.(\ref{kZ-def}). Inserting Eq.(\ref{Phi2}) into Eqs.(\ref{bc-1bis})-(\ref{bc-2bis}), one can obtain the link between the fields in the outer Rashba-dominated regions
\begin{align}\label{PsiR-PsiL-M}
    \Vector{\hat{\xi}_{\uparrow}(L/2)\\ \hat{\xi}_{\downarrow}(L/2)\\ \hat{\eta}_{\uparrow}(L/2)\\ \hat{\eta}_{\downarrow}(L/2)} = \mathsf{M}_E \Vector{\hat{\xi}_{\uparrow}(-L/2)\\ \hat{\xi}_{\downarrow}(-L/2)\\ \hat{\eta}_{\uparrow}(-L/2)\\ \hat{\eta}_{\downarrow}(-L/2)}
\end{align}
where  the  transfer matrix $\mathsf{M}_E$ depends on the energy~$E$ and on the size $L$ of the central region through two dimensionless parameters $k_Z L$ and $k_{SO} L$. Details about the derivation  of $\mathsf{M}_E$  can be found in the Appendix~\ref{AppD}. As an illustrative example, here we shall focus on the midgap value ($E=0$), which in fact well represents the entire low energy range $|E|\ll E_Z$. Moreover, since in the deep Rashba-dominated regime  $k_Z L \ll k_{SO} L$, one can keep $k_{SO} L$ finite and consider $k_Z L$ as a small parameter, performing  an expansion of $\mathsf{M}_{E=0}$ in its powers. Neglecting orders $O((k_Z L)^4)$ one obtains
\begin{widetext}
\begin{equation}
    \mathsf{M}_0 \simeq \begin{pmatrix}
    i k_{SO} L /2 & A & (1 - i k_{SO} L /2) e^{i k_{SO} L} & B\\
    A^* & -i k_{SO} L /2 & B^* & (1 + i k_{SO} L /2) e^{-i k_{SO} L}\\
    (1 + i k_{SO} L /2) e^{-i k_{SO} L} & -B & -i k_{SO} L /2 & A^* e^{2 i L k_{SO}}\\
    -B^* & (1 - i k_{SO} L /2 ) e^{i k_{SO} L} & A e^{-2 i   k_{SO} L} & i k_{SO} L /2
    \end{pmatrix} \label{transfer-Matrix0-pre}
\end{equation}
\end{widetext}
where 
\begin{align}
  A &= i \frac{-6 +  k_{SO} L (k_{SO}L +6 i)}{12   k_{SO} L}e^{2 i L k_{SO}}   (k_Z L)^2\\
B &= -i \frac{(k_{SO} L)^2+6}{12   k_{SO}L}e^{i  k_{SO}L} (k_Z L)^2 \quad.
\end{align}
The 8 entries of the transfer matrix (\ref{transfer-Matrix0-pre}) containing $A$ and $B$ couple   spin-$\uparrow$ to spin-$\downarrow$ components. Notably,  such terms are  of the order $O((k_Z L)^2)$ and  in the regime $k_Z L \ll1$  can be neglected with respect to the other terms, which are $O(1)$ with respect to the variable $k_Z L$. Then, the transfer matrix reduces to
\begin{widetext}
\begin{equation}
     \mathsf{M}_0 \simeq \begin{pmatrix}
    i\frac{k_{SO} L }{2} & 0 & (1 - i\frac{k_{SO} L }{2}) e^{i k_{SO} L} & 0\\
    0 & -i\frac{k_{SO} L }{2} & 0 & (1 + i\frac{k_{SO} L }{2}) e^{-i k_{SO} L}\\
    (1 + i\frac{k_{SO} L }{2}) e^{-i k_{SO} L} & 0 & -i\frac{k_{SO} L }{2} & 0\\
    0 & (1 - i\frac{k_{SO} L }{2}) e^{i k_{SO} L} & 0 & i\frac{k_{SO} L }{2}
    \end{pmatrix} \label{transfer-Matrix0}\quad.
\end{equation} 
\end{widetext}
The expression (\ref{transfer-Matrix0}) has precisely the form   Eq.(\ref{M-ris}) of the transfer matrix of the massless+massive Dirac model described in Sec.\ref{sec-3}, when setting $\beta_\uparrow=-\beta_\downarrow=k_{SO}L/2$, $\chi_\uparrow=-\chi_\downarrow=k_{SO}L$ and $\gamma_\uparrow=\gamma_\downarrow=\nu_\uparrow=\nu_\downarrow=0$.
Thus, in the regime $k_Z L \ll 1$, where the central region is much shorter than the Zeeman wavelength $l_Z=k_Z^{-1}$ characterizing the wavefunction (\ref{Phi2}) at $E=0$,    the transfer matrix is diagonal in spin and becomes independent of the Zeeman energy $E_Z$. Yet,  $\mathsf{M}_0$ couples massless to massive modes and still depends on $k_{SO} L$. This  parameter, which represents the ratio of the crossover region $L$ to the spin-orbit length $l_{SO}=k_{SO}^{-1}$, may be finite because of the deep Rashba-dominated regime $k_Z \ll k_{SO}$. 

\begin{figure}[h!]
  \includegraphics[width=\linewidth]{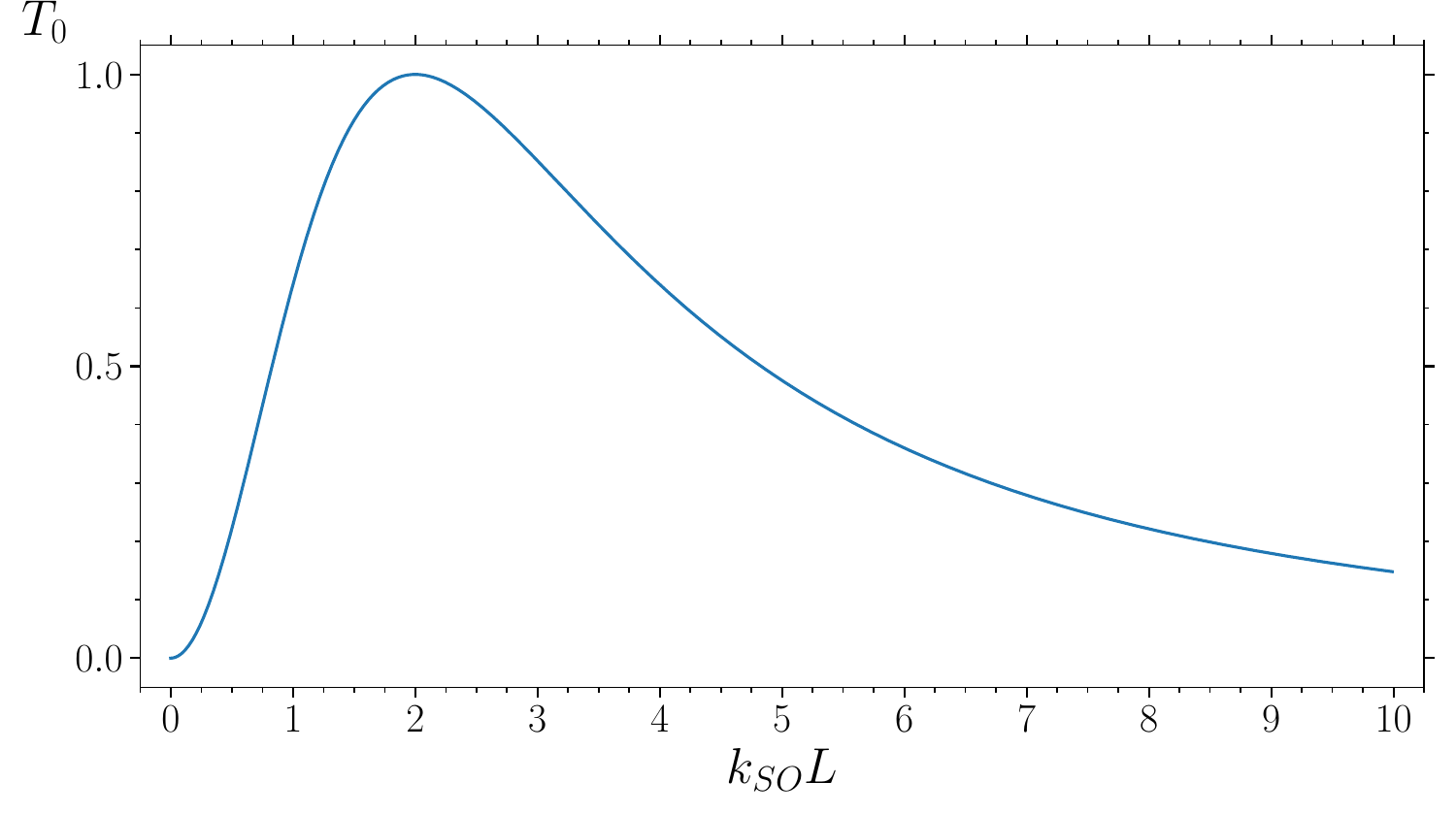}
  \caption{The transmission coefficient (\ref{T0}), plotted as a function of $k_{SO} L$, covers the entire range $T_0 \in [0,1]$.}
  \label{Fig6}
\end{figure}

In turn, the  transmission coefficient related to Eq.(\ref{transfer-Matrix0}) can be obtained from the general formula Eq.(\ref{T0-ris}), 
\begin{equation}\label{T0}
    T_0 = \frac{(k_{SO}  L)^2 }{\left(1 + (k_{SO}  L/2)^2 \right)^2} \quad,
\end{equation}
and varies over the entire range $T_0 \in [0,1]$  as a function of $k_{SO} L$, as shown  in Fig.\ref{Fig6}. In particular, while for small  values   $k_{SO}L \ll 1$ the transmission is low, $T_0 \sim  (k_{SO} L)^2$, for finite values of $k_{SO} L$  we observe from Fig.\ref{Fig6} that $T_0$ increases, and a perfect transmission $T_0 =1$ is obtained   for $k_{SO}L=2$. Then, for large values of $k_{SO} L$ the transmission decreases again as 
 $T_0 \sim 16/(k_{SO} L)^2$. The ratio of the spin-orbit length $l_{SO}=k_{SO}^{-1}$ to the distance $L$ is thus the parameter controlling the value of $T_0$.\\

\subsection{Transmission coefficient in the case of InSb}
For definiteness, we consider  here an implementation with a ballistic InSb NW   with effective electron mass $m^* = 0.015 m_e$. Two different portions of the NW  are supposed to be gated by  differently biased metals inducing opposite RSOC values, as previously sketched in Fig.\ref{Fig5}, and are separated by a crossover region   $L = 100 {\rm nm}$ where the RSOC is negligible.
In   Fig.\ref{Fig7}(a) the solid curves display the midgap  transmission coefficient $T_0=T_{E=0}$ as a function of the spin-orbit energy $E_{SO}$,   for different values of the Zeeman energy $E_Z$, obtained from the numerically exact solution of the model (\ref{H(x)_inhomo}) with the profile~(\ref{alpha-profile}) (see Appendix~\ref{AppD} for technical details).  Moreover the dashed curve describes the analytical result (\ref{T0}) obtained from the low-energy limit in the Rashba-dominated regime of the outer regions, i.e. the massless+massive Dirac model. As one can see, for $E_{SO} \rightarrow 0$, the exact transmission coefficient tends to 1, regardless of the value of $E_Z$, since all three regions become equal in such a limit.  However, for each Zeeman energy value, when $E_{SO}$ is sufficiently large to enter the deep Rashba-dominated regime ($2E_{SO}\gg E_Z$), all solid curves are well reproduced by the low-energy limit Eq.(\ref{T0})  (dashed curve), which is independent of $E_Z$. This is thus the regime where the NW gap states are helical and the setup realizes the Dirac paradox. Despite the absence of a spin-active interface, the transmission coefficient is non-vanishing because   the propagating massless modes are coupled to the evanescent massive modes.  In   Fig.\ref{Fig7}(b)   the same quantities as in panel (a) are shown, with a zoom in the range of  spin-orbit energy values up to $E_{SO} = 0.5 {\rm meV}$, which is the   realistic range presently reachable. Correspondingly,   the  range of Zeeman energy values   $E_Z$  ensuring a deep Rashba dominated regime for the external gated regions is  $E_Z < 0.1 {\rm meV}$. This implies that the linear conductance $G_0$, straightforwardly connected to the transmission coefficient through the relation  $G_0=({\rm e}^2/h) T_{E=0}$,    is   tunable   from low to high values with varying the spin-orbit energy, which can electrically be done through the gate voltage.
\begin{figure}[h!]
  \includegraphics[width=\linewidth]{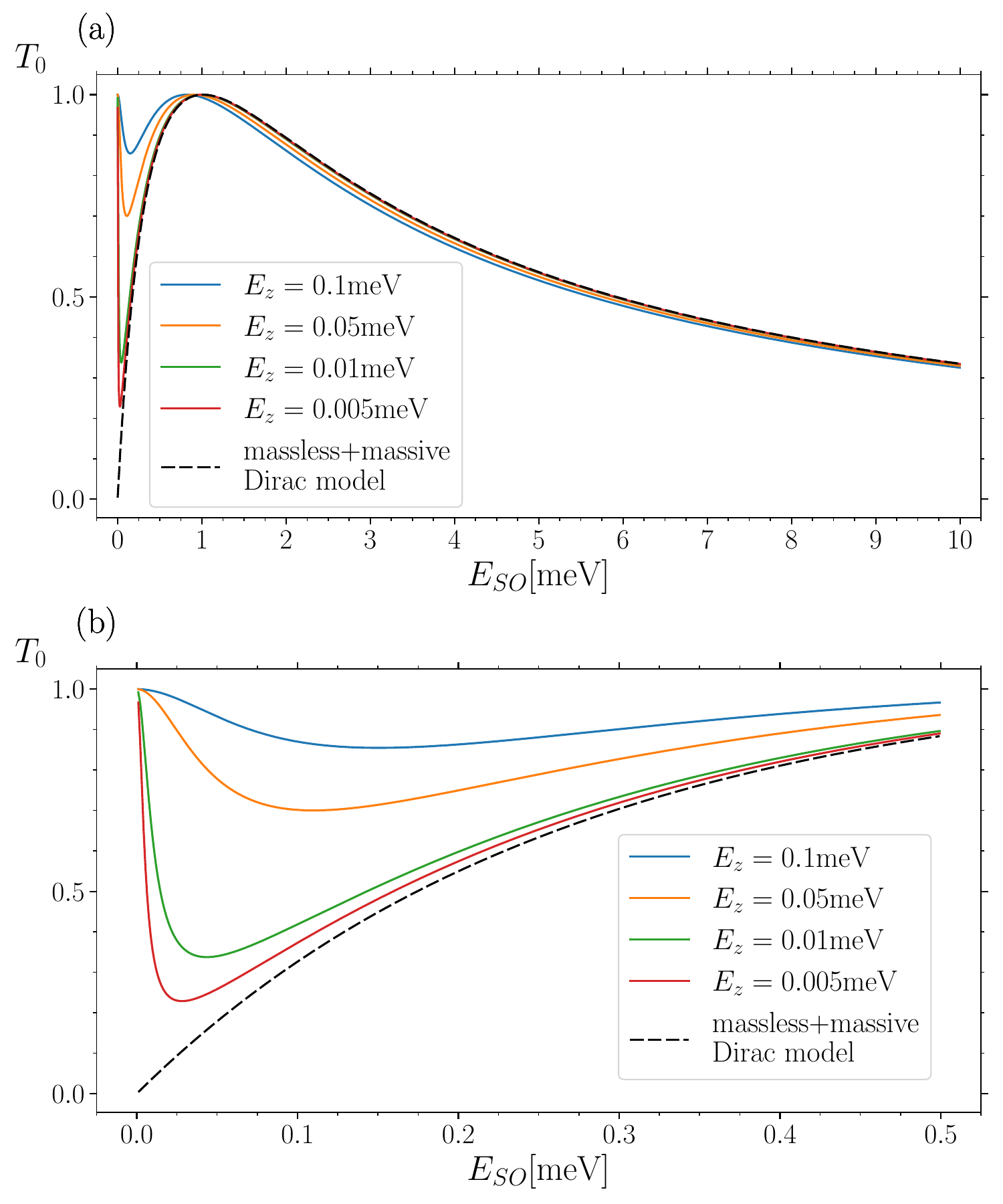}
  \caption{\label{Fig7} The Dirac paradox configuration realized with a InSb NW setup where two outer gated regions are characterized by opposite RSOC and the central region has a width $L=100\,{\rm nm}$ (see Fig.\ref{Fig5}).  The midgap transmission coefficient ($E = 0$), obtained from the numerically exact solution of the model (\ref{H(x)_inhomo}) with the profile~(\ref{alpha-profile}),   is plotted as a function of the spin orbit energy, for different values of the external magnetic field $E_Z$ (solid curves). When the Rashba-dominated regime ($2E_{SO} \gg E_Z$) is reached, the various solid curves all tend to the   dashed curve describing the result Eq.(\ref{T0}), obtained in the low energy limit from the effective massless+massive Dirac model. Panel (b) is a zoom of panel (a) in the regime of spin-orbit values that are realistic with present gating techniques.}
  \label{Fig7}
\end{figure}

Note that the electric field due to the gate is perpendicular to the nanowire axis, and the current flowing along the NW is the response to the difference in the electrochemical potentials of the two reservoirs connected to the NW. In the quantum ballistic regime considered here, the case of an electric field longitudinal to the NW axis could be realized by exposing the NW to an external electrical radiation field.
 Although such analysis goes far beyond the scope of the present paper, we mention here that, as far as the NW helical states are concerned, such problem is similar to the one studied for the helical states in Quantum Spin Hall systems. In that case, a photocurrent can be generated by a suitably localized electric pulse, and signatures of  chiral anomaly due to the helical states can be found in the chemical and temperature dependence of the spin-polarized photo-excited wavepackets.\cite{dolcini_2016}. A similar scenario can thus be expected for NWs as well.

\section{Conclusions}
\label{sec-5}
In this paper we have analyzed the Dirac paradox, illustrated in Fig.\ref{Fig1}, where an electron impinging towards an interface can seemingly neither be transmitted nor reflected.
In particular,  we have focussed on the interesting case of  one spatial dimension. Indeed, differently from higher dimensional realizations  such as   heterojunctions between two 3D Topological Insulators  where electrons can leak along the interface surface, in 1D  electrons do not have a ``way out" to escape the paradox. We have first analyzed  models that purely involve massless modes. The first model~Eq.(\ref{Ham-model-1}), where the helicity change across the interface is accounted for by a spatially inhomogeneous velocity,  leads to conclude that the paradox has no actual solution, namely it is not possible to build up a  scattering state solution. Indeed physical solutions must necessarily involve electron injection  from both sides and are characterized by a vanishing current. In contrast, the second model Eq.(\ref{Ham-model-2}), where the   helicity  change occurs through a rotation of the spin across the interface, provides a trivial solution  to the  paradox, for  it  directly introduces  a spin-active interface,  which leads to  a perfect transmission.  

Then, we have proposed a model, Eq.(\ref{Ham-model-3}), involving both massless and massive Dirac modes [see Fig.\ref{Fig2}] and we have shown that it leads to a non-trivial solution of the Dirac paradox, even for a spin-inactive interface. This is possible because of the massive modes that, despite carrying no current  for energies inside their gap, always exhibit both spin components. Thus, a massless-massive coupling at the interface indirectly enable  an incoming massless electron impinging with spin-$\uparrow$ to get transmitted  as a massless electron with spin-$\downarrow$  (see Fig.\ref{Fig3}). Properly defined scattering state solutions thus exist, and 
 the transmission coefficient depends in general on three parameters [see Eq.(\ref{T0-ris})]. 
 
Moreover, in Sec.\ref{sec-4}, we have shown that such model can be implemented in spin-orbit coupled NWs exposed to an external magnetic field, whose midgap states are characterized by massless   modes near the Fermi points $k\sim \pm 2k_{SO}$ and massive modes near $k\sim 0$ [see Fig.\ref{Fig4}]. The massless modes are helical in the deep Rashba-dominated regime ($2E_{SO} \gg E_Z$) and their helicity is determined by the sign of the RSOC. Because the latter can be tuned by state-of-the-art gating techniques, a NW with two regions  characterized by opposite RSOC values, as shown in Fig.\ref{Fig5}, is a suitable candidate to realize the Dirac paradox configuration in one spatial dimension. We have shown that the low energy limit of such inhomogeneous NW model   is precisely a particular case of the proposed model (\ref{Ham-model-3}). The resulting  transmission coefficient Eq.(\ref{T0})   varies over the full range $T_0 \in [0,1]$ (see Fig.\ref{Fig6}) as a function of the parameter $k_{SO} L$, where $L$ is the distance between the two differently gated regions and $k_{SO}$ is the spin-orbit wavevector that is directly controlled by the RSOC [see Eq.(\ref{kSO-def})].
 Focussing on the specific case of an inhomogeneous InSb NW, we have determined from model (\ref{H(x)_inhomo}) the exact  transmission coefficient, which in general depends both on the spin-orbit and  the Zeeman energies [solid curves of Fig.\ref{Fig7}]. Whenever the Rashba-dominated regime is reached,   the setup realizes the Dirac paradox configuration. Then, the transmission coefficient is well captured by the low energy limit  result (\ref{T0})  (dashed curve of Fig.\ref{Fig7})  obtained from the proposed massless+massive Dirac model and only depends on the spin-orbit energy $E_{SO}$.  Because  $E_{SO}$ can be controlled   via the gate bias coupled to the NW, the transmission coefficient and the linear conductance are electrically tuneable. 
These results thus represent a  conceptual advance in the understanding of Dirac heterojunctions  and pave the way to fruitful applications of the helical states realized in spin-orbit coupled NWs.

\acknowledgments
Discussions with Bj\"orn Trauzettel, Christoph Fleckenstein and Niccol\`o Traverso Ziani in the early stage of this work are acknowledged.

\appendix
\section{Derivation of the transfer matrix Eq.(\ref{M-ris})}
\label{AppA}
In this Appendix we provide details about the derivation of the transfer matrix (\ref{M-ris}), i.e. the most general $4 \times 4$ matrix fulfilling the requirements (\ref{cond1-M}) and (\ref{cond2-M}).
We first observe that  the former requirement (\ref{cond1-M})  straightforwardly stems from Eq.(\ref{cond1-UR}) and the property $(U_R^\dagger)^{-1}=(U_R^{-1})^\dagger$, which imply that  $U_R^{-1}$ fulfills Eq.(\ref{cond1-UR}) as well. When taking into account Eq.(\ref{cond1-UL}) and the definition (\ref{M-def}), the condition Eq.(\ref{cond1-M}) follows. Second, we observe that  the  requirement (\ref{cond2-M}) can equivalently be formulated by requiring that $\mathsf{M}$ must only involve the combinations  $\sigma_\uparrow =(\sigma_0 + \sigma_z)/2$ and $\sigma_\downarrow =(\sigma_0 - \sigma_z)/2$, i.e. $\mathsf{M}$ must have the form
\begin{equation} \label{M_definition}
\mathsf{M}=\mathsf{M}_{\uparrow} \sigma_{\uparrow} + \mathsf{M}^{}_{\downarrow} \sigma_{\downarrow}\quad,
\end{equation}
where $\mathsf{M}_{\uparrow,\downarrow}$ are $2 \times 2$ matrices acting on the massless-massive pseudospin space and   fulfilling
\begin{equation}\label{cond-Msigma}
\mathsf{M}_{\sigma}^\dagger \tau_z  \mathsf{M}^{}_{\sigma}= - \tau_z \hspace{1cm} \sigma=\uparrow,\downarrow
\end{equation}
as a consequence of  Eq.(\ref{cond1-M}) and of the properties $\sigma^2_{\uparrow,\downarrow}=\sigma_{\uparrow,\downarrow}$ and $\sigma_\uparrow \sigma_\downarrow = [\sigma_{\uparrow},\sigma_z]=[\sigma_{\downarrow},\sigma_z]=0$. 
For each spin sector $\sigma=\uparrow, \downarrow$, the requirement Eq.(\ref{cond-Msigma}) imposed on a generic 
 $2 \times 2$ complex matrix 
\begin{equation}
\mathsf{M}_{\sigma}=
\begin{pmatrix}
    a_{\sigma}       & b_{\sigma} \\
   c_{\sigma} & d_{\sigma}
\end{pmatrix}
\end{equation}
 implies  that
$|c_{\sigma}|^2-|a_{\sigma}|^2=1$, $|b_{\sigma}|^2-|d_{\sigma}|^2=1$ and $a^*_{\sigma} b_{\sigma} = c^*_{\sigma} d_{\sigma}$.
These conditions straightforwardly imply  the following expression
\begin{equation} \label{Msigma_final-form}
\mathsf{M}_{\sigma}= e^{i \nu_{\sigma}} \left( 
\begin{array}{cc} 
i \beta_{\sigma}  e^{-i \gamma_{\sigma}}  &  (1-i \beta_{\sigma}) \,e^{i \chi_{\sigma}} \\
 (1+i \beta_{\sigma}) \, e^{-i \chi_{\sigma}}&  -i\beta_{\sigma} e^{i \gamma_{\sigma}}
\end{array}
\right) \quad,
\end{equation}
which also fulfills the properties $\mathsf{M}^{-1}_{\sigma}(\beta_\sigma,\chi_\sigma,\nu_\sigma,\gamma_\sigma)=\mathsf{M}_{\sigma}(\beta_\sigma,\chi_\sigma,-\nu_\sigma,-\gamma_\sigma)$ and ${\rm det}(\mathsf{M}_\sigma)=-\exp[2i \nu_\sigma]$.
Inserting the two independent matrices $\mathsf{M}_{\uparrow}$ and $\mathsf{M}_{\downarrow}$ given in Eq.(\ref{Msigma_final-form}) into Eq.(\ref{M_definition}), the transfer matrix  $\mathsf{M}$ in the $\tau \otimes \sigma$ basis  takes the form given in Eq.(\ref{M-ris}).
 
Finally, an explicit expression can be given for $U_L$ and $U_R$ as well. The   requirement  (\ref{cond1-UL}) can always be fulfilled by choosing for $U_L$ the form
\begin{equation}
\label{UL}
U_L=\tau_0 \sigma_0  \quad.
\end{equation}
Then, the expression for $U_R=\mathsf{M}^{-1}$ following from Eq.(\ref{M-def}) can straightforwardly be obtained from Eq.(\ref{M-ris}) by exploiting the property  $\mathsf{M}^{-1}(\boldsymbol{\beta},\boldsymbol{\chi},\boldsymbol{\nu},\boldsymbol{\gamma})=\mathsf{M}(\boldsymbol{\beta},\boldsymbol{\chi},-\boldsymbol{\nu},-\boldsymbol{\gamma})$, where each bold symbols denotes the pair of related parameters, e.g. $\boldsymbol{\beta}=(\beta_\uparrow,\beta_\downarrow)$.\\


\section{details about the NW Hamiltonian and its low energy limit}
\label{AppB}
For the sake of completeness, we provide here some details about the NW Hamiltonian described in Sec.\ref{sec-4A}. Denoting by   $\Omega$  the total NW length and  re-expressing the field  in terms of its Fourier modes  $\hat{C}_k =  (\hat{c}_{k  \uparrow},\ \hat{c}_{k  \downarrow})^T$
\begin{equation} \left( \begin{array}{c} 
\hat{\Phi}_\uparrow(x)\\
\hat{\Phi}_\downarrow(x)
\end{array}
\right)=\frac{1}{\sqrt{\Omega}} \sum_k e^{i k x}  \left( \begin{array}{c} \hat{c}_{k  \uparrow} \\ \hat{c}_{k  \downarrow} \end{array} \right) \quad, \label{Phi-FT}
\end{equation}  
the NW Hamiltonian $\hat{\mathcal{H}}_{NW}$ is compactly rewritten in terms of a $2 \times 2$ matrix $H_{NW}(k)$, i.e. $\hat{\mathcal{H}}_{NW}=\sum_k \hat{C}_k^\dagger H_{NW}(k) \hat{C}_k^{}$.  In turn, this also highlights the energy scales involved in the problem. 
In particular, the first two terms acquire the form
\begin{eqnarray}
   \lefteqn{\hat{\mathcal{H}}_{kin}+\hat{\mathcal{H}}_{R}  = } & &   \label{eq:H_rk} \\
&=&    \sum_k
    \hat{C}_k^\dagger \left( \frac{\hbar^2}{2m^*}(k\sigma_0 - s_\alpha k_{SO} \sigma_z)^2 - E_{SO} \sigma_0 \right) \hat{C}_k \nonumber
    \end{eqnarray}
and describe two parabolic spin bands that are lowered by the spin-orbit energy  (\ref{ESO-def})
and  horizontally shifted by the spin-orbit wavevector (\ref{kSO-def})
with the sign (\ref{sign-alpha-def}) of the RSOC determining whether the shift   is positive or negative in $k$-axis.
Assuming $h_x>0$ for  definiteness, the Zeeman term is rewritten as
\begin{eqnarray}
\label{eq:H_z} H_{Z}  = - E_Z \sum_k \hat{C}_k^\dagger \sigma_x \hat{C}_k\quad,
\end{eqnarray}
where $E_Z$ is the Zeeman energy given  in Eq.(\ref{EZ-def}).
Summing up Eqs.(\ref{eq:H_rk}) and (\ref{eq:H_z}) the diagonalization of the resulting $H_{NW}(k)$ is straightforward. 
 Denoting $\varepsilon^0_k=\hbar^2 k^2/2 m^*$, the spectrum  consists of two energy bands  
\begin{equation}\label{NW_homo_spectrum}
E_\pm(k)=  \, \varepsilon^0_k\pm \sqrt{E_Z^2+\alpha^2 k^2}\quad,
\end{equation}
separated at $k=0$ by a  gap $2E_Z$ centered around the midgap   energy $E=0$.
The eigenfunctions related to the spectrum (\ref{NW_homo_spectrum})  are
$\psi_{k,\pm}(x)=w_{k, \pm} \exp[i k x]/\sqrt{\Omega}$. They describe plane waves with spinors
\begin{eqnarray}
w_{k,-}= \left(
\begin{array}{c}
\cos \frac{\theta_k}{2} \\  \\ 
 \sin \frac{\theta_k}{2}\,  
\end{array}\right)\hspace{0.5cm} 
w_{k, +}= \left(
\begin{array}{c}
-  \sin \frac{\theta_k}{2}\\  \\ 
\cos \frac{\theta_k}{2}
\end{array}\right)\,, \hspace{0.5cm} 
\label{wpm}
\end{eqnarray}
whose spin orientation $\mathbf{n}(k) \equiv \left( \sin\theta_k \,,0 \,,\, \cos\theta_k \right)$ lies  on the $xz$-plane and depends on the wavevector $k$, forming with the $z$-axis an angle~$\theta_k \in [0 , \pi ]$  defined through
\begin{equation}
\left\{\begin{array}{lcl}
\cos \theta_k  &=&\displaystyle  \frac{\alpha k}{\sqrt{E_Z^2+\alpha^2 k^2}}\\  
\sin \theta_k  &=&\displaystyle \frac{{E_Z}}{\sqrt{E_Z^2+\alpha^2 k^2}}
\end{array}\right.   \quad.
\end{equation}

Furthermore, for energies $|E|<E_Z$, the model also exhibits evanescent wave solutions
$\tilde{\psi}_{\kappa,\pm}(x)=\tilde{w}_{\kappa, \pm} \exp[\kappa x]/\sqrt{\Omega}$. They describe plane waves with spinors
{\small
\begin{eqnarray}
\tilde{w}_{\kappa,\pm}=\frac{1}{\sqrt{2}} \left(
\begin{array}{c}
\mp \exp\left[\pm i \mbox{arctan}\left( \alpha \kappa/\sqrt{E_Z^2-(\alpha \kappa)^2} \right) \right]  \\  \\ 
1
\end{array}\right)\hspace{0.5cm} 
\label{wtildepm}
\end{eqnarray}
}
\noindent with energy $E_\pm=-\varepsilon^0_\kappa\pm \sqrt{E_Z^2-(\alpha \kappa)^2}$. While these solutions are not normalizable in a homogeneous NW, they must be taken into account in the inhomogeneous RSOC problem.\\

 Let us now focus on the regime ($|E|\ll E_Z \ll 2 E_{SO}$) and derive an effective low energy NW Hamiltonian. 

{\it Expansion near $k =\pm 2 k_{SO}$.}  In the   deep Rashba-dominated   regime ($E_Z \ll 2 E_{SO}$),  one finds that, up to $O\big((E_Z/2 E_{SO})^2\big)$,  
\begin{eqnarray}
&E_{-}(k) \approx 0 \, \Leftrightarrow \, k \approx \pm 2k_{SO} \\
&\cos \theta_{k=\pm2k_{SO}} \approx \pm s_\alpha \quad,
\end{eqnarray}
so that the spinors (\ref{wpm}) of the lower band propagating modes   near $k \sim \pm 2 k_{SO}$  reduce  to   eigenstates  of $\sigma_z$, $(1,0)^T$ or $(0,1)^T$, depending on the sign $s_\alpha$ of the RSOC [see Eq.(\ref{sign-alpha-def})]. To extract the low energy Hamiltonian governing their dynamics, let us consider, for instance, $\alpha>0$ like in Fig.\ref{Fig4},  and focus e.g. on the vicinity of  the right Fermi point   $+2k_{SO}$. Setting $k=2k_{SO}+q$ and performing an expansion of Eqs.(\ref{eq:H_rk}) and (\ref{eq:H_z}) for $|q| \ll k_{SO}$, one obtains
\begin{eqnarray}
\lefteqn{\left. \hat{\mathcal{H}}_{NW}  \right|_{k\simeq +2k_{SO}} \simeq } & & \nonumber \\
& \simeq &   \sum_{|q|\ll k_{SO}}
    \hat{C}^\dagger_{2k_{SO}+q} \left( \begin{array}{cc} \hbar v_{SO} q & -E_Z \\ -E_Z & 8 E_{SO}\end{array} \right)   \hat{C}_{2k_{SO}+q} \nonumber \\
 & \simeq & \sum_{|q|\ll k_{SO}} \hbar v_{SO} q 	\,\hat{c}^\dagger_{2k_{SO}+q,\uparrow}  \hat{c}^{}_{2k_{SO}+q,\uparrow}  \quad,\label{near2kSO}
\end{eqnarray}
where $v_{SO} = \hbar k_{SO}/m^*$. The last line of Eq.(\ref{near2kSO}) follows from the fact that, while the spin-$\uparrow$ band is characterized by  a low-energy $\hbar v_F q$, the spin-$\downarrow$ band has a large energy $8E_{SO}$  much above the magnetic gap. The weak Zeeman energy $E_Z \ll 2E_{SO}$ cannot couple them, so that in the low energy sector $|E|\ll E_Z$  only the  spin-$\uparrow$ states matter. One can proceed in a similar manner near the $-2k_{SO}$ Fermi point, obtaining that only the  spin-$\downarrow$ states matter, proving that the states are helical. Repeating the same calculation for $\alpha<0$ one obtains the opposite helicity.
From Eq.(\ref{near2kSO}) the massless propagating low energy excitations ($|q|\ll k_{SO}$) are thus described by the set of operators 
\begin{equation}\label{xi-def}
\left\{ \begin{array}{lcl}
 \hat{\xi}_{q  \uparrow} &\doteq& \hat{c}_{2 s_\alpha k_{SO} + q, \uparrow}  \\ & & \\
 \hat{\xi}_{q  \downarrow}  &\doteq& \hat{c}_{-2 s_\alpha k_{SO} + q, \downarrow}   
\end{array}
\right.
\end{equation}
where $s_\alpha$ is given by Eq.(\ref{sign-alpha-def}).\\

{\it Expansion near $k =0$.}
In the   low energy range $|E|\ll E_Z$  there are also gapped (i.e. massive) modes, related to the upper and lower bands for $k \sim 0$ (see Fig.\ref{Fig4}). Performing an expansion of Eqs.(\ref{eq:H_rk}) and (\ref{eq:H_z}) in $q=k$ (with $|q|\ll k_{SO}$)  and introducing the new set of operators  
\begin{equation}\label{eta-def}
\left\{ \begin{array}{lcl}
\hat{\eta}_{q  \uparrow}&=    &\hat{c}_{q  \uparrow}   \\ & & \\
\hat{\eta}_{q   \downarrow} &=    &\hat{c}_{q  \downarrow}   
\end{array}
\right.\quad,
\end{equation}
one obtains the low energy expression   
\begin{align}
   \left. \hat{\mathcal{H}}_{NW} \right|_{k\simeq 0} &\simeq  - \sum_{|q|\ll k_{SO}} \hbar s_\alpha v_{SO} \, q \,
    (\hat{\eta}_{q  \uparrow}^\dagger\ \hat{\eta}_{q  \downarrow}^\dagger )    \sigma_z 
    \Vector{\hat{\eta}_{q \uparrow}\\ \hat{\eta}_{q  \downarrow}}\nonumber  \\
  -&E_Z\!\! \sum_{|q| \ll k_{SO}} \!\!
    (\hat{\eta}_{q  \uparrow}^\dagger\ \hat{\eta}_{q  \downarrow}^\dagger)
   \,  \sigma_x
     \left( \begin{array}{c} \hat{\eta}_{q  \uparrow}\\ \hat{\eta}_{q  \downarrow} \end{array} \right) \label{near0} \quad.
\end{align}
Summing up Eqs.(\ref{near2kSO}) and (\ref{near0}) one obtains a low-energy NW Hamiltonian. Moreover, one can observe that such model shares the same low-energy physics as the Dirac model given in Eq.(\ref{low-energy-model}), obtained by removing the constraints on wave vector $q$, which can therefore be regarded to as the effective low energy model for the NW. \\


\section{lattice model}
\label{AppC} 
\begin{figure}
\centering
\includegraphics[width=0.9\linewidth]{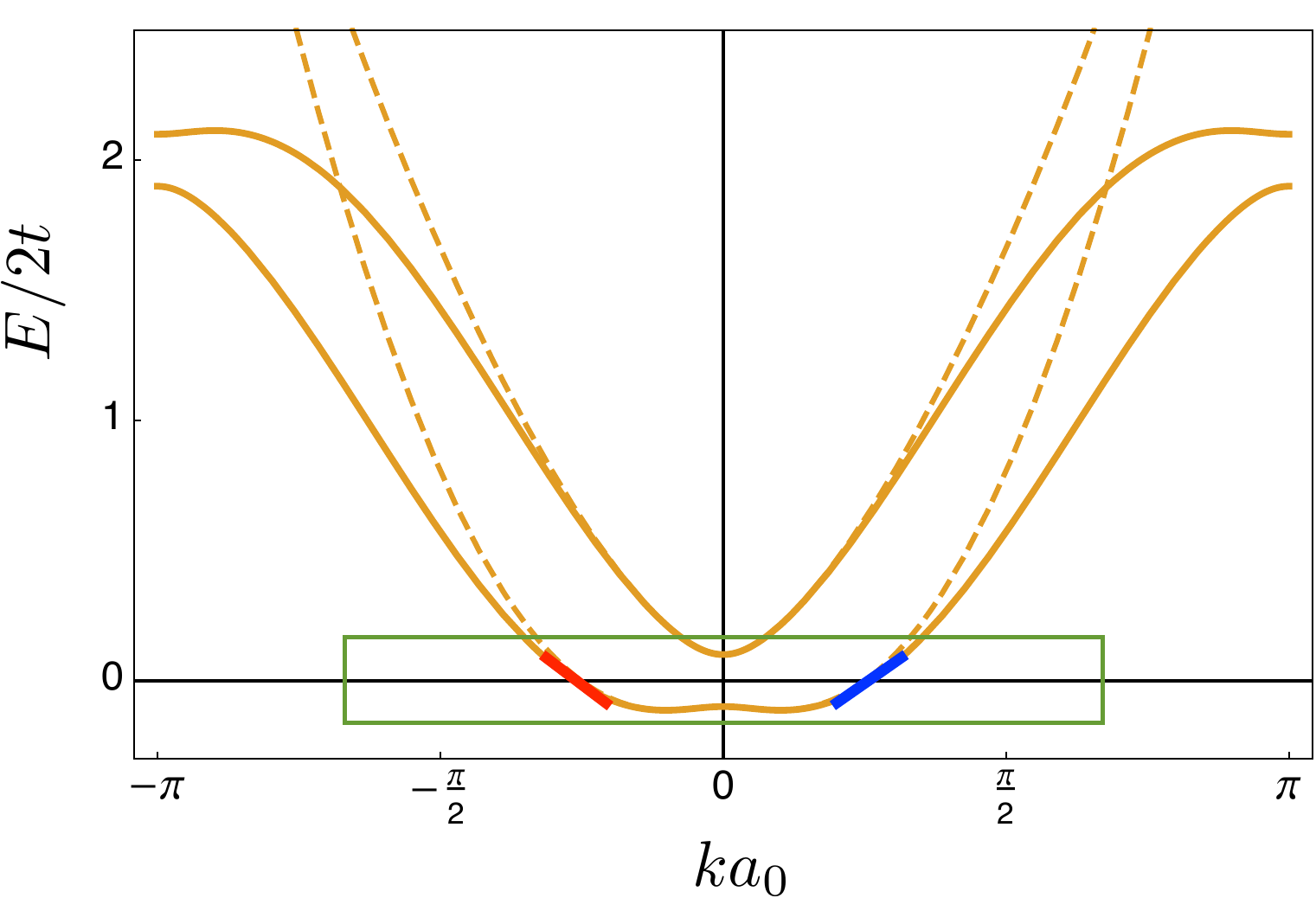} 
\caption{\label{figApp} Comparison between the  spectra of the lattice model Eq.(\ref{lattice-model-k}) (solid curves) and of the continuum model Eq. (\ref{H(x)_homo})(dashed curves). The latter captures the main features of the former in the low energy sector (green box). In particular only two helical states (red and blue thick lines) are present in the magnetic gap around $E=0$.}
\end{figure}

In this Appendix we show that the existence of one single effective massless Dirac mode, i.e. a  Weyl mode, inside the magnetic gap of the NW is not an artifact of the continuum model in Eq.(\ref{H(x)_homo}). To this purpose, we consider the following lattice model 
\begin{eqnarray}\label{lattice-model}
H&=&-t  \sum_j \left(  C_{j+1}^\dagger C_{j}  + i a \, C_{j+1}^\dagger \sigma_z C_j + b \, C_{j}^\dagger \sigma_x C_j \right.   \nonumber \\
& & \left. \hspace{2cm}- C_{j}^\dagger C_{j}\right)    + \text{H.c.}
\end{eqnarray}
where $C_{j}^\dagger = (c_{j \uparrow}^\dagger, \, c_{j \downarrow}^\dagger)$ and $c^\dagger_{j \uparrow,\downarrow}$ creates a fermion in the site $j$ with spin $\uparrow$ or $\downarrow$, respectively. Here $t$ is the nearest-neighbor hopping amplitude, while $a$ and $b$ are dimensionless parameters related to the strength of the spin-orbit coupling 
(time reversal preserving) 
and the external magnetic field  
(time reversal breaking), 
respectively. Passing to momentum space operators through $C_j=N^{-1/2} \sum_{k \in BZ} e^{i kj a_0} C_k$, where $N$ denotes the number of lattice sites, $a_0$   the lattice spacing and $k a_0 \in [-\pi,\pi]$   the lattice momentum, one gets
\begin{equation}\label{lattice-model-k}
H=2t  \sum_k C_{k}^\dagger \big\{ \!\left[1-\cos(k a_0) \right]  \sigma_0 - a \, \sin(k a_0) \sigma_z - b \, \sigma_x \big\} C_{k} \, .
\end{equation}
\noindent It is straightforward to see that Eq.(\ref{lattice-model-k}) can be considered as the lattice regularized version of the continuum model in Eq.(\ref{H(x)_homo}). Indeed in the limit $k a_0 \ll 1$, the former model  reduces to the latter upon identifying $t=\hbar^2/2 m^* a_0^2$, $a=m^* a_0 \alpha/\hbar^2$ and $b=h_x m^* a_0^2/\hbar^2$. The energy spectrum of the lattice model (\ref{lattice-model-k}) is easily obtained
\begin{equation}\label{lattice-spectrum}
E_{\pm} (k) = 2t \big[1-\cos(k) \pm \sqrt{a^2 \sin^2(k)+b^2} \big] \quad,
\end{equation}
and is plotted in Fig.\ref{figApp}   (solid lines), whereas the dashed lines display the spectrum of the continuum model (\ref{H(x)_homo}) for comparison. 
As one can see, the  low energy sector of the lattice model (green box) is  perfectly captured by the continuum theory. In particular, the bands of the full lattice model   cross the $E=0$ line in {\it two and only two points}, namely the ones   already found within the continuum model, since a gap is present at $k=0$. Thus, inside the magnetic gap, one finds only two massless helical states (red and blue thick lines), i.e. {\it one} single 1D Weyl mode.  Notably, this  is consistent with the Nielsen-Ninomiya theorem\cite{nielsen-ninomiya}, which implies, in the one dimensional case,  that  the number of left movers equals the number of right movers at any energy.  
In pass we note that, at much higher energy   (irrelevant to our purposes), a similar situation occurs: The  gap opening up at $k a_0 =\pm \pi$ leaves only two massless helical modes at $E=4t$, giving rise to  one single Dirac cone as low energy excitations around that energy. Only when time-reversal symmetry is present, i.e. for $b=0$ in Eq.(\ref{lattice-model-k}),  the two bands touch at $k=0$ and $k=\pm \pi$, where an additional Weyl mode appears.


\section{The scattering problem for the inhomogeneous NW with the profile~(\ref{alpha-profile})}
\label{AppD}
The solution of the scattering problem for the model  (\ref{H(x)_inhomo}) with the piecewise constant profile~(\ref{alpha-profile}) can be obtained from  the expression of the  electron field operator   in the three regions. For an energy $E$ within the magnetic gap ($|E|< E_Z$)  one has
\begin{widetext}
\begin{eqnarray}
\hat{\Phi}_E(x) = \left\{
\begin{array}{lcl}
\displaystyle \hat{a}_{LE}  w_{k_E,-} e^{i k_E x} + \hat{b}_{LE}  w_{-k_E,-} e^{-i k_E x} + \hat{c}_{LE} \tilde{w}_{\kappa_E,s_E} e^{\kappa_E x} & & x<-L/2\\ & & \\
\displaystyle \frac{\hat{h}_E}{\sqrt{2}} \Vector{1\\1} e^{i k_{2, E} x}+  \frac{\hat{g}_E}{\sqrt{2}}\Vector{1\\1}e^{-i k_{2, E} x} + \frac{\hat{d}_E}{\sqrt{2}}\Vector{1\\-1}e^{\kappa_{2, E} x} +   \frac{\hat{f}_E}{\sqrt{2}}\Vector{1\\-1} e^{-\kappa_{2, E} x} & & |x|<L/2 \\ & & \\
\displaystyle  \hat{a}_{RE}  w_{-k_E,-} e^{-i k_E x} +  \hat{b}_{RE}   w_{k_E,-} e^{i k_E x} + \hat{c}_{RE}    \tilde{w}_{-\kappa_E,s_E} e^{-\kappa_E x} & & x>+L/2\\
\end{array}
\right. \label{Phi-gen-3-regions}
\end{eqnarray}
\end{widetext}
where 
{\small 
\begin{eqnarray}
k_E &=&  \frac{\sqrt{2 m^*}}{\hbar}\sqrt{E+2 E_{SO}+\sqrt{4 E E_{SO}+4 E_{SO}^2+E_Z^2}} \hspace{1cm} \\
\kappa_E &=& \frac{\sqrt{2 m^*}}{\hbar}\sqrt{-E-2 E_{SO}+\sqrt{4 E E_{SO}+4 E_{SO}^2+E_Z^2}} \hspace{1cm} \\
s_E &=&\mbox{sgn}(E+E_Z^2/4 E_{SO}) \\
k_{2,E} &=& k_Z\sqrt{1+E/E_Z} \hspace{1cm} \kappa_{2,E} = k_Z\sqrt{1 -  E/E_Z}\quad,  \hspace{1cm}
\end{eqnarray}
}

\noindent while the spinors $w_{\pm k_E,-}$ and $\tilde{w}_{\pm\kappa_E,s_E}$ are given in Eqs.(\ref{wpm}) and (\ref{wtildepm}), respectively.

Imposing   the boundary conditions (\ref{bc-gen}) to the field (\ref{Phi-gen-3-regions}), one expresses the outgoing operators  $\hat{b}_{L/R\, E}$ in terms of the  operators $\hat{a}_{L E}$ and $\hat{a}_{R E}$ describing the modes  incoming from the left and from the right region, respectively. The transmission amplitudes $t_E$ and $t^\prime_E$ are then obtained through the relations $\hat{b}_{R E}=t_E \hat{a}_{L E}$ and $\hat{b}_{L  E}=t^\prime_E \hat{a}_{R E}$.
The resulting transmission coefficient $T_E=|t_E|^2=|t^\prime_E|^2$  is numerically exact and is plotted in the solid curves of Fig.\ref{Fig7} as a function of the spin orbit energy, at the midgap energy $E=0$ and for different values of the external magnetic field $E_Z$.

However, as observed in Sec.\ref{sec-4}, an analytical expression for the transmission coefficient can be obtained in the low energy limit (dashed curve in Fig.\ref{Fig7}), where the  inhomogeneous NW physics is well captured by the effective massless+massive Dirac theory. Such an expression directly follows from the 
 transfer matrix (\ref{PsiR-PsiL-M}) connecting the massless and massive fields of the outer Rashba-dominated regions, which can be obtained as follows.  
Inserting Eq.(\ref{Phi2}) into the low energy boundary conditions Eqs.(\ref{bc-1bis})-(\ref{bc-2bis}), the latter can be re-expressed in a matrix form as
\begin{align}
\mathsf{P} \Vector{\hat{\xi}_{\uparrow}(-L/2)\\ \hat{\xi}_{\downarrow}(-L/2)\\ \hat{\eta}_{\uparrow}(-L/2)\\ \eta_{\downarrow}(-L/2)} &= \mathsf{V}(-L/2) \Vector{\hat{h} \\ \hat{g} \\ \hat{d}  \\ \hat{f} } \\
\mathsf{V}(L/2) \Vector{\hat{h} \\ \hat{g} \\ \hat{d}  \\ \hat{f} } &= \mathsf{Q} \Vector{\hat{\xi}_{\uparrow}(L/2)\\ \hat{\xi}_{\downarrow}(L/2)\\ \hat{\eta}_{\uparrow}(L/2)\\ \hat{\eta}_{\downarrow}(L/2)} \quad,
\end{align}
where  the energy dependence of the operators has been dropped to make the notation lighter. Here
\begin{equation}
    \mathsf{P}  = \begin{pmatrix}
    e^{-i k_{SO} L} & 0 & 1 & 0\\
    0 & e^{i k_{SO} L} & 0 & 1\\
    k_{SO} e^{-i k_{SO} L} & 0 & -k_{SO} & 0\\
    0 & -k_{SO} e^{i k_{SO} L} & 0 & k_{SO}
    \end{pmatrix} \quad,
\end{equation}
\begin{equation}
    \mathsf{Q}  = \begin{pmatrix}
    e^{-i k_{SO} L} & 0 & 1 & 0\\
    0 & e^{i k_{SO} L} & 0 & 1\\
    -k_{SO} e^{-i k_{SO} L} & 0 & k_{SO} & 0\\
    0 & k_{SO} e^{i k_{SO} L} & 0 & - k_{SO}
    \end{pmatrix}
\end{equation}
and
{\small
\begin{eqnarray}
\lefteqn{\mathsf{V} (x) = \frac{1}{\sqrt{2}} \times } & & \label{Vmat} \\
& &\begin{pmatrix}
    e^{i k_{2, E} x} & e^{-i k_{2, E} x} & e^{\kappa_{2, E} x} & e^{-\kappa_{2, E} x}\\
    e^{i k_{2, E} x} & e^{-i k_{2, E} x} & -e^{\kappa_{2, E} x} & -e^{-\kappa_{2, E} x}\\
    i k_{2, E} e^{i k_{2, E} x} & -i k_{2, E} e^{-i k_{2, E} x} & \kappa_{2, E} e^{\kappa_{2, E} x} & -\kappa_{2, E} e^{-\kappa_{2, E} x}\\
    i k_{2, E} e^{i k_{2, E} x} & -i k_{2, E} e^{-i k_{2, E} x} & -\kappa_{2, E} e^{\kappa_{2, E} x} & \kappa_{2, E} e^{-\kappa_{2, E} x}\\
    \end{pmatrix}\nonumber
\end{eqnarray}
}

\noindent The transfer matrix $\mathsf{M}_E$ appearing in Eq.(\ref{PsiR-PsiL-M}) can thus straightforwardly be obtained as $\mathsf{M}_E = \mathsf{Q}^{-1} \mathsf{V}(L/2) \mathsf{V}^{-1}(-L/2) \mathsf{P}$. In particular, setting the energy to the midgap value $E=0$ and expanding  in the parameter $k_Z L$ one obtains Eq.(\ref{transfer-Matrix0-pre}), up to $O((k_Z L)^4)$ terms.

\end{document}